\documentclass[longauth]{aaEC}
\usepackage{graphicx}
\usepackage{natbib}
\usepackage{scalerel}
\usepackage{siunitx}
\usepackage{placeins}
\usepackage{booktabs}
\usepackage{lastpage}
\usepackage[table]{xcolor}
\usepackage{txfonts}
\usepackage{longtable}
\usepackage{graphicx}
\usepackage{comment}
\usepackage[pdfencoding=auto,psdextra]{hyperref}
\hypersetup{
    colorlinks=true,
    linkcolor=blue,
    filecolor=magenta,      
    urlcolor=blue,
    citecolor=blue
}
\usepackage[capitalise]{cleveref}
\crefname{section}{Sect.}{Sects.}
\Crefname{section}{Section}{Sections}

\makeatletter
\renewcommand*\aa@pageof{, page \thepage{} of \pageref*{LastPage}}
\makeatother

\usepackage[utf8]{inputenc}

\usepackage[switch, modulo]{lineno}

\usepackage{euclid}

\begin{document}
%
%
\title{\vspace{-0.24cm}\Euclid: Finding strong gravitational lenses in the Early Release Observations using convolutional neural networks\thanks{This paper is published on behalf of the Euclid Consortium.}}    


\newcommand{\orcid}[1]{} 
\author{B.~C.~Nagam\orcid{0000-0002-3724-7694}\thanks{\email{b.c.nagam@rug.nl}}\inst{\ref{aff1},\ref{aff2}}
\and J.~A.~Acevedo~Barroso\orcid{0000-0002-9654-1711}\inst{\ref{aff3}}
\and J.~Wilde\orcid{0000-0002-4460-7379}\inst{\ref{aff4}}
\and I.~T.~Andika\orcid{0000-0001-6102-9526}\inst{\ref{aff5},\ref{aff6}}
\and A.~Manj\'on-Garc\'ia\orcid{0000-0002-7413-8825}\inst{\ref{aff7}}
\and R.~Pearce-Casey\inst{\ref{aff8}}
\and D.~Stern\orcid{0000-0003-2686-9241}\inst{\ref{aff9}}
\and J.~W.~Nightingale\orcid{0000-0002-8987-7401}\inst{\ref{aff10}}
\and L.~A.~Moustakas\orcid{0000-0003-3030-2360}\inst{\ref{aff9}}
\and K.~McCarthy\orcid{0000-0001-6857-018X}\inst{\ref{aff9}}
\and E.~Moravec\orcid{0000-0001-9793-5416}\inst{\ref{aff11}}
\and L.~Leuzzi\orcid{0009-0006-4479-7017}\inst{\ref{aff12},\ref{aff13}}
\and K.~Rojas\orcid{0000-0003-1391-6854}\inst{\ref{aff14}}
\and S.~Serjeant\orcid{0000-0002-0517-7943}\inst{\ref{aff8}}
\and T.~E.~Collett\orcid{0000-0001-5564-3140}\inst{\ref{aff15}}
\and P.~Matavulj\orcid{0000-0003-0229-7189}\inst{\ref{aff14}}
\and M.~Walmsley\orcid{0000-0002-6408-4181}\inst{\ref{aff16},\ref{aff17}}
\and B.~Cl\'ement\orcid{0000-0002-7966-3661}\inst{\ref{aff3},\ref{aff18}}
\and C.~Tortora\orcid{0000-0001-7958-6531}\inst{\ref{aff19}}
\and R.~Gavazzi\orcid{0000-0002-5540-6935}\inst{\ref{aff20},\ref{aff21}}
\and R.~B.~Metcalf\orcid{0000-0003-3167-2574}\inst{\ref{aff12},\ref{aff13}}
\and C.~M.~O'Riordan\orcid{0000-0003-2227-1998}\inst{\ref{aff6}}
\and G.~Verdoes~Kleijn\orcid{0000-0001-5803-2580}\inst{\ref{aff1}}
\and L.~V.~E.~Koopmans\orcid{0000-0003-1840-0312}\inst{\ref{aff1}}
\and E.~A.~Valentijn\inst{\ref{aff1}}
\and V.~Busillo\orcid{0009-0000-6049-1073}\inst{\ref{aff19},\ref{aff22},\ref{aff23}}
\and S.~Schuldt\orcid{0000-0003-2497-6334}\inst{\ref{aff24},\ref{aff25}}
\and F.~Courbin\orcid{0000-0003-0758-6510}\inst{\ref{aff4},\ref{aff26}}
\and G.~Vernardos\orcid{0000-0001-8554-7248}\inst{\ref{aff27},\ref{aff28}}
\and M.~Meneghetti\orcid{0000-0003-1225-7084}\inst{\ref{aff13},\ref{aff29}}
\and A.~D\'iaz-S\'anchez\orcid{0000-0003-0748-4768}\inst{\ref{aff7}}
\and J.~M.~Diego\orcid{0000-0001-9065-3926}\inst{\ref{aff30}}
\and L.~R.~Ecker\inst{\ref{aff31},\ref{aff32}}
\and T.~T.~Thai\orcid{0000-0002-8408-4816}\inst{\ref{aff20},\ref{aff33}}
\and A.~R.~Cooray\orcid{0000-0002-3892-0190}\inst{\ref{aff34}}
\and H.~M.~Courtois\orcid{0000-0003-0509-1776}\inst{\ref{aff35}}
\and L.~Delchambre\orcid{0000-0003-2559-408X}\inst{\ref{aff36}}
\and G.~Despali\orcid{0000-0001-6150-4112}\inst{\ref{aff12},\ref{aff13},\ref{aff29}}
\and D.~Sluse\orcid{0000-0001-6116-2095}\inst{\ref{aff36}}
\and L.~Ulivi\orcid{0009-0001-3291-5382}\inst{\ref{aff37},\ref{aff38},\ref{aff39}}
\and A.~Melo\orcid{0000-0002-6449-3970}\inst{\ref{aff6},\ref{aff5}}
\and P.~Corcho-Caballero\orcid{0000-0001-6327-7080}\inst{\ref{aff1}}
\and B.~Altieri\orcid{0000-0003-3936-0284}\inst{\ref{aff40}}
\and A.~Amara\inst{\ref{aff41}}
\and S.~Andreon\orcid{0000-0002-2041-8784}\inst{\ref{aff42}}
\and N.~Auricchio\orcid{0000-0003-4444-8651}\inst{\ref{aff13}}
\and H.~Aussel\orcid{0000-0002-1371-5705}\inst{\ref{aff43}}
\and C.~Baccigalupi\orcid{0000-0002-8211-1630}\inst{\ref{aff44},\ref{aff45},\ref{aff46},\ref{aff47}}
\and M.~Baldi\orcid{0000-0003-4145-1943}\inst{\ref{aff48},\ref{aff13},\ref{aff29}}
\and A.~Balestra\orcid{0000-0002-6967-261X}\inst{\ref{aff49}}
\and S.~Bardelli\orcid{0000-0002-8900-0298}\inst{\ref{aff13}}
\and P.~Battaglia\orcid{0000-0002-7337-5909}\inst{\ref{aff13}}
\and D.~Bonino\orcid{0000-0002-3336-9977}\inst{\ref{aff50}}
\and E.~Branchini\orcid{0000-0002-0808-6908}\inst{\ref{aff51},\ref{aff52},\ref{aff42}}
\and M.~Brescia\orcid{0000-0001-9506-5680}\inst{\ref{aff22},\ref{aff19}}
\and J.~Brinchmann\orcid{0000-0003-4359-8797}\inst{\ref{aff53},\ref{aff54}}
\and A.~Caillat\inst{\ref{aff20}}
\and S.~Camera\orcid{0000-0003-3399-3574}\inst{\ref{aff55},\ref{aff56},\ref{aff50}}
\and V.~Capobianco\orcid{0000-0002-3309-7692}\inst{\ref{aff50}}
\and C.~Carbone\orcid{0000-0003-0125-3563}\inst{\ref{aff25}}
\and J.~Carretero\orcid{0000-0002-3130-0204}\inst{\ref{aff57},\ref{aff58}}
\and S.~Casas\orcid{0000-0002-4751-5138}\inst{\ref{aff59}}
\and M.~Castellano\orcid{0000-0001-9875-8263}\inst{\ref{aff60}}
\and G.~Castignani\orcid{0000-0001-6831-0687}\inst{\ref{aff13}}
\and S.~Cavuoti\orcid{0000-0002-3787-4196}\inst{\ref{aff19},\ref{aff23}}
\and A.~Cimatti\inst{\ref{aff61}}
\and C.~Colodro-Conde\inst{\ref{aff62}}
\and G.~Congedo\orcid{0000-0003-2508-0046}\inst{\ref{aff63}}
\and C.~J.~Conselice\orcid{0000-0003-1949-7638}\inst{\ref{aff17}}
\and L.~Conversi\orcid{0000-0002-6710-8476}\inst{\ref{aff64},\ref{aff40}}
\and Y.~Copin\orcid{0000-0002-5317-7518}\inst{\ref{aff65}}
\and M.~Cropper\orcid{0000-0003-4571-9468}\inst{\ref{aff66}}
\and A.~Da~Silva\orcid{0000-0002-6385-1609}\inst{\ref{aff67},\ref{aff68}}
\and H.~Degaudenzi\orcid{0000-0002-5887-6799}\inst{\ref{aff69}}
\and G.~De~Lucia\orcid{0000-0002-6220-9104}\inst{\ref{aff45}}
\and A.~M.~Di~Giorgio\orcid{0000-0002-4767-2360}\inst{\ref{aff70}}
\and J.~Dinis\orcid{0000-0001-5075-1601}\inst{\ref{aff67},\ref{aff68}}
\and F.~Dubath\orcid{0000-0002-6533-2810}\inst{\ref{aff69}}
\and C.~A.~J.~Duncan\orcid{0009-0003-3573-0791}\inst{\ref{aff17}}
\and X.~Dupac\inst{\ref{aff40}}
\and S.~Dusini\orcid{0000-0002-1128-0664}\inst{\ref{aff71}}
\and M.~Fabricius\orcid{0000-0002-7025-6058}\inst{\ref{aff32},\ref{aff31}}
\and M.~Farina\orcid{0000-0002-3089-7846}\inst{\ref{aff70}}
\and S.~Farrens\orcid{0000-0002-9594-9387}\inst{\ref{aff43}}
\and S.~Ferriol\inst{\ref{aff65}}
\and M.~Frailis\orcid{0000-0002-7400-2135}\inst{\ref{aff45}}
\and E.~Franceschi\orcid{0000-0002-0585-6591}\inst{\ref{aff13}}
\and M.~Fumana\orcid{0000-0001-6787-5950}\inst{\ref{aff25}}
\and K.~George\orcid{0000-0002-1734-8455}\inst{\ref{aff31}}
\and W.~Gillard\orcid{0000-0003-4744-9748}\inst{\ref{aff72}}
\and B.~Gillis\orcid{0000-0002-4478-1270}\inst{\ref{aff63}}
\and C.~Giocoli\orcid{0000-0002-9590-7961}\inst{\ref{aff13},\ref{aff29}}
\and P.~G\'omez-Alvarez\orcid{0000-0002-8594-5358}\inst{\ref{aff73},\ref{aff40}}
\and A.~Grazian\orcid{0000-0002-5688-0663}\inst{\ref{aff49}}
\and F.~Grupp\inst{\ref{aff32},\ref{aff31}}
\and L.~Guzzo\orcid{0000-0001-8264-5192}\inst{\ref{aff24},\ref{aff42}}
\and S.~V.~H.~Haugan\orcid{0000-0001-9648-7260}\inst{\ref{aff74}}
\and J.~Hoar\inst{\ref{aff40}}
\and W.~Holmes\inst{\ref{aff9}}
\and I.~Hook\orcid{0000-0002-2960-978X}\inst{\ref{aff75}}
\and F.~Hormuth\inst{\ref{aff76}}
\and A.~Hornstrup\orcid{0000-0002-3363-0936}\inst{\ref{aff77},\ref{aff78}}
\and P.~Hudelot\inst{\ref{aff21}}
\and K.~Jahnke\orcid{0000-0003-3804-2137}\inst{\ref{aff79}}
\and M.~Jhabvala\inst{\ref{aff80}}
\and B.~Joachimi\orcid{0000-0001-7494-1303}\inst{\ref{aff81}}
\and E.~Keih\"anen\orcid{0000-0003-1804-7715}\inst{\ref{aff82}}
\and S.~Kermiche\orcid{0000-0002-0302-5735}\inst{\ref{aff72}}
\and B.~Kubik\orcid{0009-0006-5823-4880}\inst{\ref{aff65}}
\and K.~Kuijken\orcid{0000-0002-3827-0175}\inst{\ref{aff83}}
\and M.~K\"ummel\orcid{0000-0003-2791-2117}\inst{\ref{aff31}}
\and M.~Kunz\orcid{0000-0002-3052-7394}\inst{\ref{aff84}}
\and H.~Kurki-Suonio\orcid{0000-0002-4618-3063}\inst{\ref{aff85},\ref{aff86}}
\and R.~Laureijs\inst{\ref{aff87},\ref{aff1}}
\and D.~Le~Mignant\orcid{0000-0002-5339-5515}\inst{\ref{aff20}}
\and S.~Ligori\orcid{0000-0003-4172-4606}\inst{\ref{aff50}}
\and P.~B.~Lilje\orcid{0000-0003-4324-7794}\inst{\ref{aff74}}
\and V.~Lindholm\orcid{0000-0003-2317-5471}\inst{\ref{aff85},\ref{aff86}}
\and I.~Lloro\orcid{0000-0001-5966-1434}\inst{\ref{aff88}}
\and G.~Mainetti\orcid{0000-0003-2384-2377}\inst{\ref{aff89}}
\and E.~Maiorano\orcid{0000-0003-2593-4355}\inst{\ref{aff13}}
\and O.~Mansutti\orcid{0000-0001-5758-4658}\inst{\ref{aff45}}
\and O.~Marggraf\orcid{0000-0001-7242-3852}\inst{\ref{aff90}}
\and K.~Markovic\orcid{0000-0001-6764-073X}\inst{\ref{aff9}}
\and M.~Martinelli\orcid{0000-0002-6943-7732}\inst{\ref{aff60},\ref{aff91}}
\and N.~Martinet\orcid{0000-0003-2786-7790}\inst{\ref{aff20}}
\and F.~Marulli\orcid{0000-0002-8850-0303}\inst{\ref{aff12},\ref{aff13},\ref{aff29}}
\and R.~Massey\orcid{0000-0002-6085-3780}\inst{\ref{aff92}}
\and E.~Medinaceli\orcid{0000-0002-4040-7783}\inst{\ref{aff13}}
\and M.~Melchior\inst{\ref{aff14}}
\and Y.~Mellier\inst{\ref{aff93},\ref{aff21}}
\and E.~Merlin\orcid{0000-0001-6870-8900}\inst{\ref{aff60}}
\and G.~Meylan\inst{\ref{aff3}}
\and M.~Moresco\orcid{0000-0002-7616-7136}\inst{\ref{aff12},\ref{aff13}}
\and L.~Moscardini\orcid{0000-0002-3473-6716}\inst{\ref{aff12},\ref{aff13},\ref{aff29}}
\and R.~Nakajima\orcid{0009-0009-1213-7040}\inst{\ref{aff90}}
\and C.~Neissner\orcid{0000-0001-8524-4968}\inst{\ref{aff94},\ref{aff58}}
\and R.~C.~Nichol\orcid{0000-0003-0939-6518}\inst{\ref{aff41}}
\and S.-M.~Niemi\inst{\ref{aff87}}
\and C.~Padilla\orcid{0000-0001-7951-0166}\inst{\ref{aff94}}
\and S.~Paltani\orcid{0000-0002-8108-9179}\inst{\ref{aff69}}
\and F.~Pasian\orcid{0000-0002-4869-3227}\inst{\ref{aff45}}
\and K.~Pedersen\inst{\ref{aff95}}
\and W.~J.~Percival\orcid{0000-0002-0644-5727}\inst{\ref{aff96},\ref{aff97},\ref{aff98}}
\and V.~Pettorino\inst{\ref{aff87}}
\and S.~Pires\orcid{0000-0002-0249-2104}\inst{\ref{aff43}}
\and G.~Polenta\orcid{0000-0003-4067-9196}\inst{\ref{aff99}}
\and M.~Poncet\inst{\ref{aff100}}
\and L.~A.~Popa\inst{\ref{aff101}}
\and L.~Pozzetti\orcid{0000-0001-7085-0412}\inst{\ref{aff13}}
\and F.~Raison\orcid{0000-0002-7819-6918}\inst{\ref{aff32}}
\and R.~Rebolo\inst{\ref{aff62},\ref{aff102},\ref{aff103}}
\and A.~Renzi\orcid{0000-0001-9856-1970}\inst{\ref{aff104},\ref{aff71}}
\and J.~Rhodes\orcid{0000-0002-4485-8549}\inst{\ref{aff9}}
\and G.~Riccio\inst{\ref{aff19}}
\and E.~Romelli\orcid{0000-0003-3069-9222}\inst{\ref{aff45}}
\and M.~Roncarelli\orcid{0000-0001-9587-7822}\inst{\ref{aff13}}
\and E.~Rossetti\orcid{0000-0003-0238-4047}\inst{\ref{aff48}}
\and R.~Saglia\orcid{0000-0003-0378-7032}\inst{\ref{aff31},\ref{aff32}}
\and Z.~Sakr\orcid{0000-0002-4823-3757}\inst{\ref{aff105},\ref{aff106},\ref{aff107}}
\and A.~G.~S\'anchez\orcid{0000-0003-1198-831X}\inst{\ref{aff32}}
\and D.~Sapone\orcid{0000-0001-7089-4503}\inst{\ref{aff108}}
\and B.~Sartoris\orcid{0000-0003-1337-5269}\inst{\ref{aff31},\ref{aff45}}
\and M.~Schirmer\orcid{0000-0003-2568-9994}\inst{\ref{aff79}}
\and P.~Schneider\orcid{0000-0001-8561-2679}\inst{\ref{aff90}}
\and T.~Schrabback\orcid{0000-0002-6987-7834}\inst{\ref{aff109}}
\and A.~Secroun\orcid{0000-0003-0505-3710}\inst{\ref{aff72}}
\and G.~Seidel\orcid{0000-0003-2907-353X}\inst{\ref{aff79}}
\and S.~Serrano\orcid{0000-0002-0211-2861}\inst{\ref{aff110},\ref{aff111},\ref{aff112}}
\and C.~Sirignano\orcid{0000-0002-0995-7146}\inst{\ref{aff104},\ref{aff71}}
\and G.~Sirri\orcid{0000-0003-2626-2853}\inst{\ref{aff29}}
\and J.~Skottfelt\orcid{0000-0003-1310-8283}\inst{\ref{aff113}}
\and L.~Stanco\orcid{0000-0002-9706-5104}\inst{\ref{aff71}}
\and J.-L.~Starck\orcid{0000-0003-2177-7794}\inst{\ref{aff43}}
\and J.~Steinwagner\orcid{0000-0001-7443-1047}\inst{\ref{aff32}}
\and P.~Tallada-Cresp\'{i}\orcid{0000-0002-1336-8328}\inst{\ref{aff57},\ref{aff58}}
\and D.~Tavagnacco\orcid{0000-0001-7475-9894}\inst{\ref{aff45}}
\and A.~N.~Taylor\inst{\ref{aff63}}
\and H.~I.~Teplitz\orcid{0000-0002-7064-5424}\inst{\ref{aff114}}
\and I.~Tereno\inst{\ref{aff67},\ref{aff115}}
\and R.~Toledo-Moreo\orcid{0000-0002-2997-4859}\inst{\ref{aff116}}
\and F.~Torradeflot\orcid{0000-0003-1160-1517}\inst{\ref{aff58},\ref{aff57}}
\and A.~Tsyganov\inst{\ref{aff117}}
\and I.~Tutusaus\orcid{0000-0002-3199-0399}\inst{\ref{aff106}}
\and L.~Valenziano\orcid{0000-0002-1170-0104}\inst{\ref{aff13},\ref{aff118}}
\and T.~Vassallo\orcid{0000-0001-6512-6358}\inst{\ref{aff31},\ref{aff45}}
\and A.~Veropalumbo\orcid{0000-0003-2387-1194}\inst{\ref{aff42},\ref{aff52},\ref{aff51}}
\and Y.~Wang\orcid{0000-0002-4749-2984}\inst{\ref{aff114}}
\and J.~Weller\orcid{0000-0002-8282-2010}\inst{\ref{aff31},\ref{aff32}}
\and A.~Zacchei\orcid{0000-0003-0396-1192}\inst{\ref{aff45},\ref{aff44}}
\and E.~Zucca\orcid{0000-0002-5845-8132}\inst{\ref{aff13}}
\and C.~Burigana\orcid{0000-0002-3005-5796}\inst{\ref{aff119},\ref{aff118}}
\and A.~Mora\orcid{0000-0002-1922-8529}\inst{\ref{aff120}}
\and M.~P\"ontinen\orcid{0000-0001-5442-2530}\inst{\ref{aff85}}
\and V.~Scottez\inst{\ref{aff93},\ref{aff121}}}
										   
\institute{Kapteyn Astronomical Institute, University of Groningen, PO Box 800, 9700 AV Groningen, The Netherlands\label{aff1}
\and
Minnesota Institute for Astrophysics, University of Minnesota, 116 Church St SE, Minneapolis, MN 55455, USA\label{aff2}
\and
Institute of Physics, Laboratory of Astrophysics, Ecole Polytechnique F\'ed\'erale de Lausanne (EPFL), Observatoire de Sauverny, 1290 Versoix, Switzerland\label{aff3}
\and
Institut de Ci\`{e}ncies del Cosmos (ICCUB), Universitat de Barcelona (IEEC-UB), Mart\'{i} i Franqu\`{e}s 1, 08028 Barcelona, Spain\label{aff4}
\and
Technical University of Munich, TUM School of Natural Sciences, Physics Department, James-Franck-Str.~1, 85748 Garching, Germany\label{aff5}
\and
Max-Planck-Institut f\"ur Astrophysik, Karl-Schwarzschild-Str.~1, 85748 Garching, Germany\label{aff6}
\and
Departamento F\'isica Aplicada, Universidad Polit\'ecnica de Cartagena, Campus Muralla del Mar, 30202 Cartagena, Murcia, Spain\label{aff7}
\and
School of Physical Sciences, The Open University, Milton Keynes, MK7 6AA, UK\label{aff8}
\and
Jet Propulsion Laboratory, California Institute of Technology, 4800 Oak Grove Drive, Pasadena, CA, 91109, USA\label{aff9}
\and
School of Mathematics, Statistics and Physics, Newcastle University, Herschel Building, Newcastle-upon-Tyne, NE1 7RU, UK\label{aff10}
\and
Green Bank Observatory, P.O. Box 2, Green Bank, WV 24944, USA\label{aff11}
\and
Dipartimento di Fisica e Astronomia "Augusto Righi" - Alma Mater Studiorum Universit\`a di Bologna, via Piero Gobetti 93/2, 40129 Bologna, Italy\label{aff12}
\and
INAF-Osservatorio di Astrofisica e Scienza dello Spazio di Bologna, Via Piero Gobetti 93/3, 40129 Bologna, Italy\label{aff13}
\and
University of Applied Sciences and Arts of Northwestern Switzerland, School of Engineering, 5210 Windisch, Switzerland\label{aff14}
\and
Institute of Cosmology and Gravitation, University of Portsmouth, Portsmouth PO1 3FX, UK\label{aff15}
\and
David A. Dunlap Department of Astronomy \& Astrophysics, University of Toronto, 50 St George Street, Toronto, Ontario M5S 3H4, Canada\label{aff16}
\and
Jodrell Bank Centre for Astrophysics, Department of Physics and Astronomy, University of Manchester, Oxford Road, Manchester M13 9PL, UK\label{aff17}
\and
SCITAS, Ecole Polytechnique F\'ed\'erale de Lausanne (EPFL), 1015 Lausanne, Switzerland\label{aff18}
\and
INAF-Osservatorio Astronomico di Capodimonte, Via Moiariello 16, 80131 Napoli, Italy\label{aff19}
\and
Aix-Marseille Universit\'e, CNRS, CNES, LAM, Marseille, France\label{aff20}
\and
Institut d'Astrophysique de Paris, UMR 7095, CNRS, and Sorbonne Universit\'e, 98 bis boulevard Arago, 75014 Paris, France\label{aff21}
\and
Department of Physics "E. Pancini", University Federico II, Via Cinthia 6, 80126, Napoli, Italy\label{aff22}
\and
INFN section of Naples, Via Cinthia 6, 80126, Napoli, Italy\label{aff23}
\and
Dipartimento di Fisica "Aldo Pontremoli", Universit\`a degli Studi di Milano, Via Celoria 16, 20133 Milano, Italy\label{aff24}
\and
INAF-IASF Milano, Via Alfonso Corti 12, 20133 Milano, Italy\label{aff25}
\and
Instituci\'o Catalana de Recerca i Estudis Avan\c{c}ats (ICREA), Passeig de Llu\'{\i}s Companys 23, 08010 Barcelona, Spain\label{aff26}
\and
Department of Physics and Astronomy, Lehman College of the CUNY, Bronx, NY 10468, USA\label{aff27}
\and
American Museum of Natural History, Department of Astrophysics, New York, NY 10024, USA\label{aff28}
\and
INFN-Sezione di Bologna, Viale Berti Pichat 6/2, 40127 Bologna, Italy\label{aff29}
\and
Instituto de F\'isica de Cantabria, Edificio Juan Jord\'a, Avenida de los Castros, 39005 Santander, Spain\label{aff30}
\and
Universit\"ats-Sternwarte M\"unchen, Fakult\"at f\"ur Physik, Ludwig-Maximilians-Universit\"at M\"unchen, Scheinerstrasse 1, 81679 M\"unchen, Germany\label{aff31}
\and
Max Planck Institute for Extraterrestrial Physics, Giessenbachstr. 1, 85748 Garching, Germany\label{aff32}
\and
National Astronomical Observatory of Japan, 2-21-1 Osawa, Mitaka, Tokyo 181-8588, Japan\label{aff33}
\and
Department of Physics \& Astronomy, University of California Irvine, Irvine CA 92697, USA\label{aff34}
\and
UCB Lyon 1, CNRS/IN2P3, IUF, IP2I Lyon, 4 rue Enrico Fermi, 69622 Villeurbanne, France\label{aff35}
\and
STAR Institute, Quartier Agora - All\'ee du six Ao\^ut, 19c B-4000 Li\`ege, Belgium\label{aff36}
\and
University of Trento, Via Sommarive 14, I-38123 Trento, Italy\label{aff37}
\and
Dipartimento di Fisica e Astronomia, Universit\`{a} di Firenze, via G. Sansone 1, 50019 Sesto Fiorentino, Firenze, Italy\label{aff38}
\and
INAF-Osservatorio Astrofisico di Arcetri, Largo E. Fermi 5, 50125, Firenze, Italy\label{aff39}
\and
ESAC/ESA, Camino Bajo del Castillo, s/n., Urb. Villafranca del Castillo, 28692 Villanueva de la Ca\~nada, Madrid, Spain\label{aff40}
\and
School of Mathematics and Physics, University of Surrey, Guildford, Surrey, GU2 7XH, UK\label{aff41}
\and
INAF-Osservatorio Astronomico di Brera, Via Brera 28, 20122 Milano, Italy\label{aff42}
\and
Universit\'e Paris-Saclay, Universit\'e Paris Cit\'e, CEA, CNRS, AIM, 91191, Gif-sur-Yvette, France\label{aff43}
\and
IFPU, Institute for Fundamental Physics of the Universe, via Beirut 2, 34151 Trieste, Italy\label{aff44}
\and
INAF-Osservatorio Astronomico di Trieste, Via G. B. Tiepolo 11, 34143 Trieste, Italy\label{aff45}
\and
INFN, Sezione di Trieste, Via Valerio 2, 34127 Trieste TS, Italy\label{aff46}
\and
SISSA, International School for Advanced Studies, Via Bonomea 265, 34136 Trieste TS, Italy\label{aff47}
\and
Dipartimento di Fisica e Astronomia, Universit\`a di Bologna, Via Gobetti 93/2, 40129 Bologna, Italy\label{aff48}
\and
INAF-Osservatorio Astronomico di Padova, Via dell'Osservatorio 5, 35122 Padova, Italy\label{aff49}
\and
INAF-Osservatorio Astrofisico di Torino, Via Osservatorio 20, 10025 Pino Torinese (TO), Italy\label{aff50}
\and
Dipartimento di Fisica, Universit\`a di Genova, Via Dodecaneso 33, 16146, Genova, Italy\label{aff51}
\and
INFN-Sezione di Genova, Via Dodecaneso 33, 16146, Genova, Italy\label{aff52}
\and
Instituto de Astrof\'isica e Ci\^encias do Espa\c{c}o, Universidade do Porto, CAUP, Rua das Estrelas, PT4150-762 Porto, Portugal\label{aff53}
\and
Faculdade de Ci\^encias da Universidade do Porto, Rua do Campo de Alegre, 4150-007 Porto, Portugal\label{aff54}
\and
Dipartimento di Fisica, Universit\`a degli Studi di Torino, Via P. Giuria 1, 10125 Torino, Italy\label{aff55}
\and
INFN-Sezione di Torino, Via P. Giuria 1, 10125 Torino, Italy\label{aff56}
\and
Centro de Investigaciones Energ\'eticas, Medioambientales y Tecnol\'ogicas (CIEMAT), Avenida Complutense 40, 28040 Madrid, Spain\label{aff57}
\and
Port d'Informaci\'{o} Cient\'{i}fica, Campus UAB, C. Albareda s/n, 08193 Bellaterra (Barcelona), Spain\label{aff58}
\and
Institute for Theoretical Particle Physics and Cosmology (TTK), RWTH Aachen University, 52056 Aachen, Germany\label{aff59}
\and
INAF-Osservatorio Astronomico di Roma, Via Frascati 33, 00078 Monteporzio Catone, Italy\label{aff60}
\and
Dipartimento di Fisica e Astronomia "Augusto Righi" - Alma Mater Studiorum Universit\`a di Bologna, Viale Berti Pichat 6/2, 40127 Bologna, Italy\label{aff61}
\and
Instituto de Astrof\'{\i}sica de Canarias, V\'{\i}a L\'actea, 38205 La Laguna, Tenerife, Spain\label{aff62}
\and
Institute for Astronomy, University of Edinburgh, Royal Observatory, Blackford Hill, Edinburgh EH9 3HJ, UK\label{aff63}
\and
European Space Agency/ESRIN, Largo Galileo Galilei 1, 00044 Frascati, Roma, Italy\label{aff64}
\and
Universit\'e Claude Bernard Lyon 1, CNRS/IN2P3, IP2I Lyon, UMR 5822, Villeurbanne, F-69100, France\label{aff65}
\and
Mullard Space Science Laboratory, University College London, Holmbury St Mary, Dorking, Surrey RH5 6NT, UK\label{aff66}
\and
Departamento de F\'isica, Faculdade de Ci\^encias, Universidade de Lisboa, Edif\'icio C8, Campo Grande, PT1749-016 Lisboa, Portugal\label{aff67}
\and
Instituto de Astrof\'isica e Ci\^encias do Espa\c{c}o, Faculdade de Ci\^encias, Universidade de Lisboa, Campo Grande, 1749-016 Lisboa, Portugal\label{aff68}
\and
Department of Astronomy, University of Geneva, ch. d'Ecogia 16, 1290 Versoix, Switzerland\label{aff69}
\and
INAF-Istituto di Astrofisica e Planetologia Spaziali, via del Fosso del Cavaliere, 100, 00100 Roma, Italy\label{aff70}
\and
INFN-Padova, Via Marzolo 8, 35131 Padova, Italy\label{aff71}
\and
Aix-Marseille Universit\'e, CNRS/IN2P3, CPPM, Marseille, France\label{aff72}
\and
FRACTAL S.L.N.E., calle Tulip\'an 2, Portal 13 1A, 28231, Las Rozas de Madrid, Spain\label{aff73}
\and
Institute of Theoretical Astrophysics, University of Oslo, P.O. Box 1029 Blindern, 0315 Oslo, Norway\label{aff74}
\and
Department of Physics, Lancaster University, Lancaster, LA1 4YB, UK\label{aff75}
\and
Felix Hormuth Engineering, Goethestr. 17, 69181 Leimen, Germany\label{aff76}
\and
Technical University of Denmark, Elektrovej 327, 2800 Kgs. Lyngby, Denmark\label{aff77}
\and
Cosmic Dawn Center (DAWN), Denmark\label{aff78}
\and
Max-Planck-Institut f\"ur Astronomie, K\"onigstuhl 17, 69117 Heidelberg, Germany\label{aff79}
\and
NASA Goddard Space Flight Center, Greenbelt, MD 20771, USA\label{aff80}
\and
Department of Physics and Astronomy, University College London, Gower Street, London WC1E 6BT, UK\label{aff81}
\and
Department of Physics and Helsinki Institute of Physics, Gustaf H\"allstr\"omin katu 2, 00014 University of Helsinki, Finland\label{aff82}
\and
Leiden Observatory, Leiden University, Einsteinweg 55, 2333 CC Leiden, The Netherlands\label{aff83}
\and
Universit\'e de Gen\`eve, D\'epartement de Physique Th\'eorique and Centre for Astroparticle Physics, 24 quai Ernest-Ansermet, CH-1211 Gen\`eve 4, Switzerland\label{aff84}
\and
Department of Physics, P.O. Box 64, 00014 University of Helsinki, Finland\label{aff85}
\and
Helsinki Institute of Physics, Gustaf H{\"a}llstr{\"o}min katu 2, University of Helsinki, Helsinki, Finland\label{aff86}
\and
European Space Agency/ESTEC, Keplerlaan 1, 2201 AZ Noordwijk, The Netherlands\label{aff87}
\and
NOVA optical infrared instrumentation group at ASTRON, Oude Hoogeveensedijk 4, 7991PD, Dwingeloo, The Netherlands\label{aff88}
\and
Centre de Calcul de l'IN2P3/CNRS, 21 avenue Pierre de Coubertin 69627 Villeurbanne Cedex, France\label{aff89}
\and
Universit\"at Bonn, Argelander-Institut f\"ur Astronomie, Auf dem H\"ugel 71, 53121 Bonn, Germany\label{aff90}
\and
INFN-Sezione di Roma, Piazzale Aldo Moro, 2 - c/o Dipartimento di Fisica, Edificio G. Marconi, 00185 Roma, Italy\label{aff91}
\and
Department of Physics, Institute for Computational Cosmology, Durham University, South Road, Durham, DH1 3LE, UK\label{aff92}
\and
Institut d'Astrophysique de Paris, 98bis Boulevard Arago, 75014, Paris, France\label{aff93}
\and
Institut de F\'{i}sica d'Altes Energies (IFAE), The Barcelona Institute of Science and Technology, Campus UAB, 08193 Bellaterra (Barcelona), Spain\label{aff94}
\and
DARK, Niels Bohr Institute, University of Copenhagen, Jagtvej 155, 2200 Copenhagen, Denmark\label{aff95}
\and
Waterloo Centre for Astrophysics, University of Waterloo, Waterloo, Ontario N2L 3G1, Canada\label{aff96}
\and
Department of Physics and Astronomy, University of Waterloo, Waterloo, Ontario N2L 3G1, Canada\label{aff97}
\and
Perimeter Institute for Theoretical Physics, Waterloo, Ontario N2L 2Y5, Canada\label{aff98}
\and
Space Science Data Center, Italian Space Agency, via del Politecnico snc, 00133 Roma, Italy\label{aff99}
\and
Centre National d'Etudes Spatiales -- Centre spatial de Toulouse, 18 avenue Edouard Belin, 31401 Toulouse Cedex 9, France\label{aff100}
\and
Institute of Space Science, Str. Atomistilor, nr. 409 M\u{a}gurele, Ilfov, 077125, Romania\label{aff101}
\and
Consejo Superior de Investigaciones Cientificas, Calle Serrano 117, 28006 Madrid, Spain\label{aff102}
\and
Universidad de La Laguna, Departamento de Astrof\'{\i}sica, 38206 La Laguna, Tenerife, Spain\label{aff103}
\and
Dipartimento di Fisica e Astronomia "G. Galilei", Universit\`a di Padova, Via Marzolo 8, 35131 Padova, Italy\label{aff104}
\and
Institut f\"ur Theoretische Physik, University of Heidelberg, Philosophenweg 16, 69120 Heidelberg, Germany\label{aff105}
\and
Institut de Recherche en Astrophysique et Plan\'etologie (IRAP), Universit\'e de Toulouse, CNRS, UPS, CNES, 14 Av. Edouard Belin, 31400 Toulouse, France\label{aff106}
\and
Universit\'e St Joseph; Faculty of Sciences, Beirut, Lebanon\label{aff107}
\and
Departamento de F\'isica, FCFM, Universidad de Chile, Blanco Encalada 2008, Santiago, Chile\label{aff108}
\and
Universit\"at Innsbruck, Institut f\"ur Astro- und Teilchenphysik, Technikerstr. 25/8, 6020 Innsbruck, Austria\label{aff109}
\and
Institut d'Estudis Espacials de Catalunya (IEEC),  Edifici RDIT, Campus UPC, 08860 Castelldefels, Barcelona, Spain\label{aff110}
\and
Satlantis, University Science Park, Sede Bld 48940, Leioa-Bilbao, Spain\label{aff111}
\and
Institute of Space Sciences (ICE, CSIC), Campus UAB, Carrer de Can Magrans, s/n, 08193 Barcelona, Spain\label{aff112}
\and
Centre for Electronic Imaging, Open University, Walton Hall, Milton Keynes, MK7~6AA, UK\label{aff113}
\and
Infrared Processing and Analysis Center, California Institute of Technology, Pasadena, CA 91125, USA\label{aff114}
\and
Instituto de Astrof\'isica e Ci\^encias do Espa\c{c}o, Faculdade de Ci\^encias, Universidade de Lisboa, Tapada da Ajuda, 1349-018 Lisboa, Portugal\label{aff115}
\and
Universidad Polit\'ecnica de Cartagena, Departamento de Electr\'onica y Tecnolog\'ia de Computadoras,  Plaza del Hospital 1, 30202 Cartagena, Spain\label{aff116}
\and
Centre for Information Technology, University of Groningen, P.O. Box 11044, 9700 CA Groningen, The Netherlands\label{aff117}
\and
INFN-Bologna, Via Irnerio 46, 40126 Bologna, Italy\label{aff118}
\and
INAF, Istituto di Radioastronomia, Via Piero Gobetti 101, 40129 Bologna, Italy\label{aff119}
\and
Aurora Technology for European Space Agency (ESA), Camino bajo del Castillo, s/n, Urbanizacion Villafranca del Castillo, Villanueva de la Ca\~nada, 28692 Madrid, Spain\label{aff120}
\and
ICL, Junia, Universit\'e Catholique de Lille, LITL, 59000 Lille, France\label{aff121}}       

%
%
 \abstract{
The Early Release Observations (ERO) from \Euclid have detected several new galaxy-galaxy strong gravitational lenses, with the all-sky survey expected to find \num{170000} new systems, greatly enhancing studies of dark matter, dark energy, and constraints on the cosmological parameters. As a first step, visual inspection of all galaxies in one of the ERO fields (Perseus) was carried out to identify candidate strong lensing systems and compared to the predictions from Convolutional Neural Networks (CNNs). However, the entire ERO data set is too large for expert visual inspection. In this paper, we therefore extend the CNN analysis to the whole ERO data set, using different CNN architectures and methodologies. 
Using five CNN architectures, we identified 8,469 strong gravitational lens candidates from \IE-band cutouts of 13 \Euclid ERO fields, narrowing them to 97 through visual inspection, including 14 grade A and 31 grade B candidates. We present the spectroscopic confirmation of a strong gravitational lensing candidate, EUCL\,J081705.61$+$702348.8. The foreground lensing galaxy, an early-type system at $z=0.335$, and the background source, a star-forming galaxy at $z=1.475$ with [\ion{O}{II}] emission, are both identified. Lens modelling using the Euclid strong lens modelling pipeline reveals two distinct arcs in a lensing configuration, with an Einstein radius of $\ang{;;1.18} \pm \ang{;;0.03}$, confirming the lensing nature of the system. These findings highlight the importance of a broad CNN search to efficiently reduce candidates, followed by visual inspection to eliminate false positives and achieve a high-purity sample of strong lenses in \Euclid.

}
%
%
%
   \titlerunning{\Euclid\/: Finding strong lenses in the ERO using CNNs}
   \authorrunning{B. C. Nagam et al.}
   
   \maketitle
%
%
%
%
   
\section{\label{sc:Intro}Introduction}

The deflection of light rays from a distant background source by a massive foreground object, a phenomenon known as strong gravitational lensing, produces multiple resolved images, arcs, or Einstein rings, depending upon relative positions and alignment of the source, lens, and observer. Strong lensing has many important applications, such as (i) mapping the mass distribution of galaxy \citep{mass_distribution_0, mass_distribution_1, 2004astro.ph.12596K,sgl_app_2019, SGL_mass_distribution_turyshev}, (ii) providing constraints on dark energy \citep{Dark_Energy_5, Dark_Energy_3, Dark_Energy_4, Dark_Energy_2, Dark_Energy_6, Dark_Energy_7} and dark matter \citep{Tortora_2010, SGL_DM_2, SGL_DM_2021, Vegetti2024}, (iii) constraining the slope of inner mass density profile (e.g., \citealp{IMD_2002}; \citealp{IMD_2001}; \citealp{IMD_2007}; \citealp{ISD_2009}; \citealp{IMD_2012}; \citealp{ICD_2015b}; \citealp{ISD_2015}; \citealp{IMD_2020}; \citealp{2022MNRAS.516..336S}), and, (iv) measurement of the Hubble constant ($H_{0}$) using time-delay cosmography between multiple imaged sources \citep{SGL_H_1990,SGL_H_2003, SGL_H_2018,SGL_H_Treu_2022, SGL_time_delay_anowar, birrer2024time}.

Gravitational lensing is sensitive to the presence of all foreground matter, regardless of whether this matter is in the form of visible baryonic matter or dark matter. To a very good approximation, dark matter appears to be collisionless, and lacking in detectable signatures of interactions via any of the fundamental forces except gravity. As such, gravitational lensing is one of the few probes of the dynamics and physical properties of dark matter particles that drive the growth of large-scale structures in the Universe and dominate the cosmic matter budget.

The \Euclid mission \citep{EuclidSkyOverview} is a 1.2\,m space survey telescope with primary science objectives including measurements of cosmic shear from weak lensing, and galaxy clustering, from which cosmological parameter constraints follow. \Euclid was successfully launched in July 2023 and is in the process of observing approximately one-third of the sky in its wide survey \citep{Scaramella-EP1}, and $\sim$$50\,\deg^2$ in its deep fields, about two magnitudes deeper than the wide survey. \Euclid's instruments consist of a visible imager (VIS), a near-infrared imager and a slitless spectrometer (NISP), and are described in more detail in \citet{EuclidSkyVIS} and \citet{EuclidSkyNISP}, respectively. The vast imaging and spectroscopic dataset from the \Euclid survey enables a wide range of legacy science. Data sets such as these generally excel in their potential for discovering rare objects. 


\Euclid is widely expected to be transformative in the field of strong gravitational lensing, with $\num{170000}$ galaxy-scale strong lenses predicted in its wide survey \citep[e.g.,][]{100000_lenses_3}, increasing the number of known strong lenses by about two orders of magnitude. This creates a new frontier of rare object discovery, such as the detection of highly magnified high-redshift (e.g., $z>2$) background sources, or of ``Jackpot'' systems with two background galaxies magnified at different redshifts being magnified by the same foreground lens. Each of these opens the possibility of a range of new legacy science goals for \Euclid, including the direct detection and statistical characterization of dark matter halo substructure \citep[e.g.,][]{Vegetti+14,Hezaveh+16,Li+16,Li+17,Despali+18, 2023MNRAS.521.2342O}, geometrical constraints on the dark energy equation of state \citep[e.g.,][]{jackpot, Daniel_Stern_2010,SHLIVKO2024138826}, and 
calibration of initial mass function of early-type galaxies \citep{Treu_2010_main, Barnabe_main, Leier_main}.

However, identifying these \mbox{$\sim$$10^5$} strong lensing systems among the \mbox{$\sim$$10^9$} \Euclid galaxies is a non-trivial task. Driven by \Euclid and other major sky surveys, many authors have developed machine learning methods for finding strong gravitational lensing systems \citep[e.g.,][]{petrillo17,Petrillo_2,Petrillo_3,lanusse2018cmu,SGL_CNN_app_pour,schaefer2018deep,SGL_CNN_app_2019b,SGL_CNN_app_2019c,SGL_CNN_app_2020,SGL_CNN_app_2021,SGL_CNN_app_christ,canameras2020holismokes,SGL_CNN_app_christ,huang2017densely,Rezaei,Andika_2023,Fabrizio,Savary,Rojas,wilde2022detecting,nagam,nagam2}.

The first data released from \Euclid were the Early Release Observations (ERO) \citep{EROData,EROOrion,EROGalGCs,ERONearbyGals,EROFornaxGCs,EROPerseusOverview,EROPerseusDGs,EROPerseusICL,EROLensData,EROLensVISdropouts}. As a first test for the existence of a large population of strong gravitational lenses, \citet[][hereafter AB24]{AcevedoBarroso24} inspected all galaxies in the Perseus ERO field \citep{cuillandre_Perseus}, finding three gravitational lensing systems (termed grade A candidates), 13 probable systems (grade B candidates), and 52 possible strong lenses (grade C candidates). These were compared to the results of 21 CNN architectures in \citet[][hereafter PC24]{ERO_lens_search_CNN}, with the result that, at best, CNNs produce lens candidate lists that are only $\sim$$10$\% pure. The best performing CNN was also one of the deepest networks, pre-trained on over a hundred million galaxy classifications from the Galaxy Zoo citizen science project (\citealp{lintott_zoo}), with a final layer fine-tuned by re-training on strong lenses painted onto real galaxy images from the \Euclid ERO data sets. However, some similarities were observed in the false positives, such as the presence of spiral arms in the target galaxy, suggesting that further improvements to the CNN architecture or training process are possible.

The entire \Euclid ERO data set is too large for expert volunteers to conduct such an examination of every galaxy in a timely fashion. Therefore, in this paper, we present the results of our strong gravitational lens search in the remainder of the \Euclid ERO data, using a CNN-based approach followed by visual inspection. In performing this analysis, we take an approach inspired by previous works \citep{jacobs2017finding,jacobs2019finding,Andika_2023,nagam}, by comparing the outputs of various CNN lens finders. Inevitably, the vast majority of \Euclid's strong gravitational lenses will first be identified through an initial CNN search, followed by visual inspection. However, this CNN search inevitably misses some strong lensing systems, resulting in a population of false negatives. Here, we attempt to mitigate this by considering a diversity of CNN approaches, searching for systems that are detected by one CNN and not another. This approach is particularly important for discovering rare lensing configurations that will contribute to \Euclid's legacy science goals.

This paper is structured as follows. In \cref{sec:data}, we summarise the \Euclid ERO data characteristics and the processing applied to them for the detection of strong gravitational lenses with CNNs. \Cref{sec:methods} presents the methodology used. The results are presented in \cref{sec:results}, and the results of recent spectroscopic follow-up are presented in \cref{Sec:Spectroscopy}. The summary and the conclusions are discussed in \cref{sec:discussion}.

\section{\Euclid Early Release Observation data}\label{sec:data}

The \Euclid ERO data release covers 17 galactic and extragalactic fields. The objective was to acquire scientific observations for communication and early scientific results before the start of the nominal mission \citep{EROcite,EROData}.

In this work, we focus on all the ERO fields except for the Perseus cluster, which has already been mined for lenses in AB24 and PC24. All ERO fields are covered by both the VIS and NISP instruments. However, in \cref{table:ero_statistics}, we present only the VIS statistics for brevity. We refer the reader to \citet{EROData} for a more complete overview.

Given the O($10^6$) sources in the whole ERO catalogues, we narrow our selection to extended sources brighter than 23 magnitudes in VIS \IE-band. This is in agreement with the selection used in AB24 and PC24. Thus, the selection corresponds to $\rm{\texttt{MAG\_AUTO}}<23$ and $\rm{\texttt{CLASS\_STAR}}<0.5$ in the VIS catalogues, for a total of \num{377472} sources. This classification is conservative and includes a significant fraction of stars.

We create postage stamps for every selected source. Each stamp covers $\ang{;;9.9} \times \ang{;;9.9}$, corresponding to $99 \times 99$ pixels in VIS and $33 \times 33$ pixels in the NISP bands. The stamp size is kept large enough to include any galaxy-scale lensing effects, given that the typical Einstein radius is rarely larger than $\ang{;;3}$, but also small enough to keep only the relevant information, without contamination from nearby sources.


\section{Methodology}\label{sec:methods}

\subsection{Candidate selection}

We employed the following CNN models: \texttt{4-Layer\,CNN} \citep{manjon_thesis_2021}, \texttt{DenseLens} \citep{nagam}, \texttt{Lens-CLR} (Andika et al. in prep.), \texttt{MRC-95} \citep{wilde2023applications}, and \texttt{Naberrie} \citep{wilde2023applications}. For a detailed overview of these models and their application to the \Euclid ERO Perseus field, see PC24. We applied each of these CNN models to the \num{377472} extracted VIS cutouts, obtaining five scores per source. To mitigate the impact of noise in individual classifications, we clipped scores below 0.1 and computed the geometric mean as the final score. The arbitrary clipping step helps to prevent the removal of good candidates due to noisy classifications.

Each CNN model was trained on different data sets and used different architectures, thereby capturing a wide variety of lensing and non-lensing features (see PC24 for more details). This aggregating method allowed for a more robust and consistent identification of lens candidates. From the initial 16 ERO fields, three fields (Horsehead, IC\,342, and NGC\,6254) were excluded from our final analysis since they did not contain any viable candidates. The cutouts were then ranked based on the geometric mean values, and the top 650 cutouts from each of the remaining 13 fields were selected, resulting in a total of \num{8450} cutouts chosen by the CNNs for further inspection. We adopt the same approach as AB24 and PC24, utilizing both the VIS and NISP bands for visual inspection, because the human eye is good at detecting subtle colour variations \citep{marshall2009automated}. However, we exclude the NISP cutouts from the CNN inference process to focus on a single band, which is effective at capturing morphological features \citep{Petrillo_2}, while acknowledging that incorporating additional bands could provide additional information and may improve classification performance.

\subsection{Visual inspection of candidates}
\begin{figure}[!htb]
    \centering
    \includegraphics[width=\columnwidth]{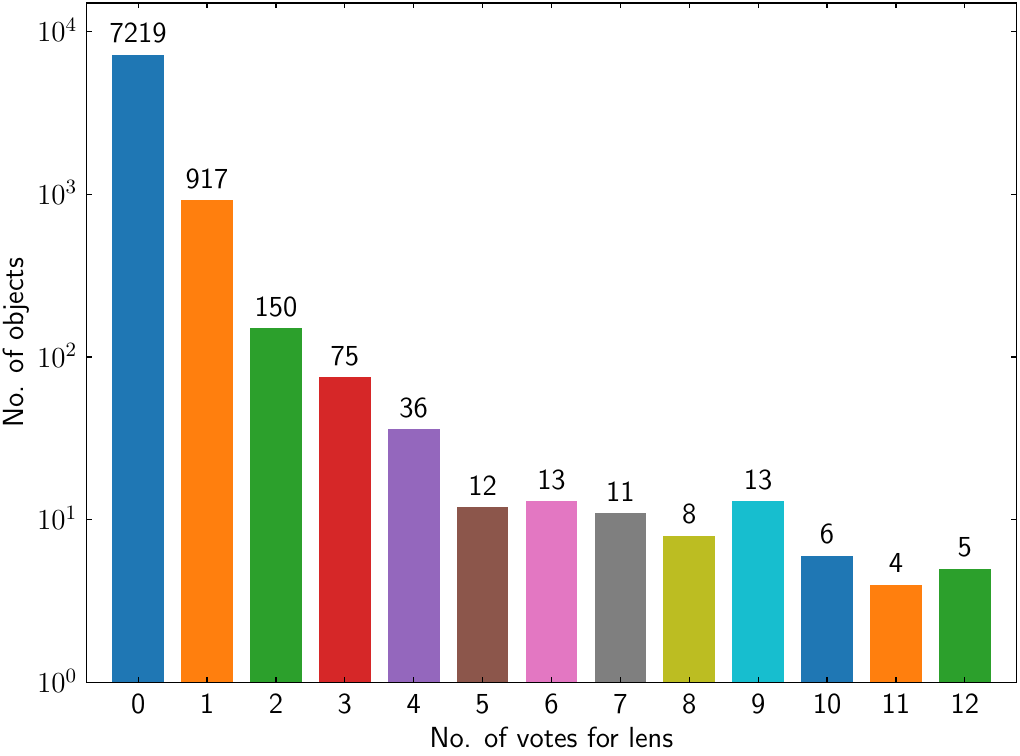} 
    \caption{The voting results for stage 1 of the visual inspection process, which involved 12 people, were based on \num{8469} candidates selected by CNNs.}
    \label{fig:stage_1_visual_inspec}
\end{figure}

During the visual analysis of the results of individual models, 19 additional notable candidates were identified and included in the selection. This brought the total number of candidates to \num{8469}, which were then subjected to a first round of visual inspection by 12 experts, in which the experts rejected obvious non-lenses. The voting results from this stage of visual inspection are illustrated in \cref{fig:stage_1_visual_inspec}, showing the distribution of votes each candidate received. Based on the voting, 183 candidates that received three or more votes were selected for further inspection.

However, during the subsequent analysis, 32 of these 183 candidates were identified as duplicates, which were removed from the sample. These duplicates arose from the VIS source catalogue itself containing multiple entries for the same physical source, with slight positional shifts within \mbox{\ang{;;1}} between the duplicate entries. This left 151 unique candidates, which then underwent a second, more rigorous round of visual inspection. This stage involved 15 experts who graded each candidate based on the presence of lensing features, using a grading scheme (see AB24) that categorized the candidates into four grades: A, B, C, and X. In short, the grading scheme implies the following:
\begin{itemize}
    \item Grade A represents definite lenses with clear lensing features;
    \item Grade B suggests the presence of lensing features, though confirmation requires additional data;
    \item Grade C indicates lensing features that could also be attributed to other physical phenomena;
    \item Grade X refers to objects that are definitively not lenses.
\end{itemize}

\section{Results}\label{sec:results}
We present here an overview of the grade A, B, and C candidates found in our search of the ERO fields. Our search methodology combined multiple CNN models with visual classification to ensure robust candidate identification.
\subsection{CNNs with visual classification}
We identified a total of 97 lens candidates, distributed as 14 grade A candidates, 31 grade B candidates, and 52 grade C candidates with visual classification. We also identified 54 objects as grade X (non-lenses) during our visual inspection process. The catalogue of all these candidates is shown in \cref{sec:table_catalog}.

\begin{figure*}[!htb]%
    \centering
    {{\includegraphics[width=\textwidth]{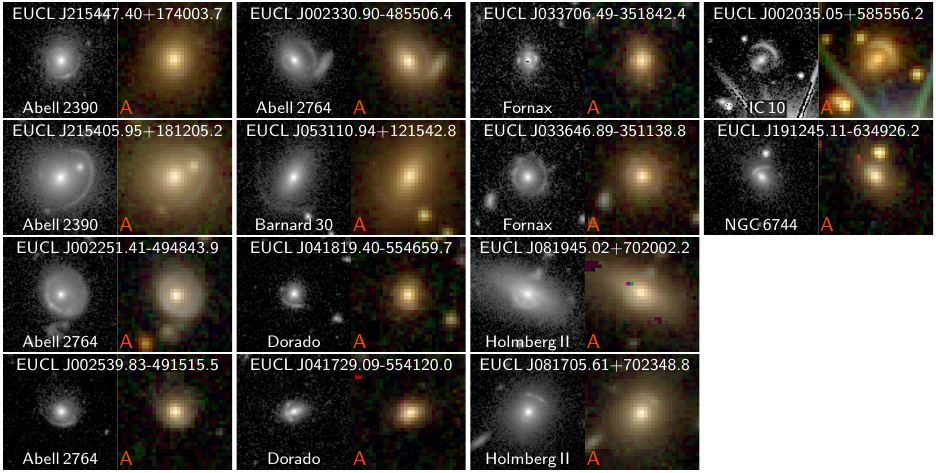}}}
    \caption{Mosaic of the grade A lens candidates from the second round of visual inspection. For each candidate, we show the high-resolution \IE-band cutout on the left and the lower-resolution \HE, \YE, \IE composite on the right. The IAU name and field name are displayed at the top and bottom of the \IE-band cutout, respectively, and the final joint grade is shown in red at the bottom of the composite cutout. Each cutout is $\ang{;;9.9} \times \ang{;;9.9}$ in size.}
    \label{fig:A_Grade}
\end{figure*}

\begin{figure*}[!htb]%
    \centering
\includegraphics[width=\textwidth]{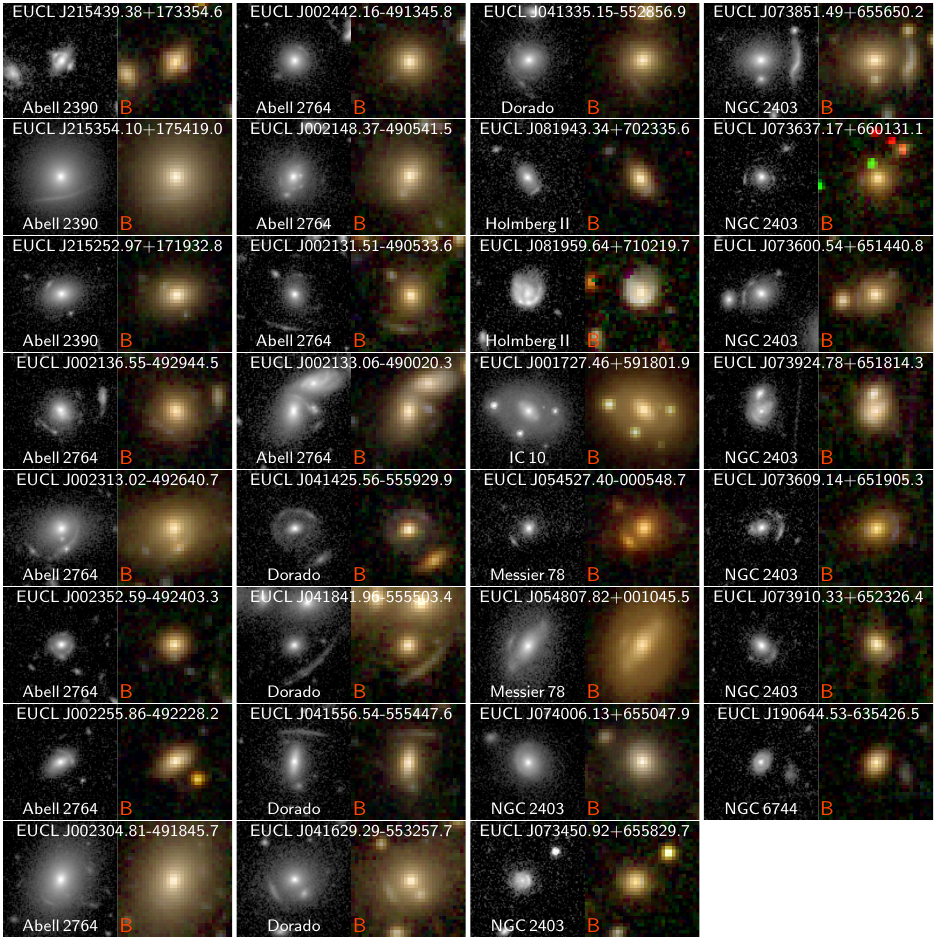}
    \caption{Mosaic of the grade B lens candidates from the second round of visual inspection. For each candidate, we show the high-resolution \IE-band cutout on the left and the lower-resolution \HE, \YE, \IE composite on the right. The IAU name and field name are displayed at the top and bottom of the \IE-band cutout, respectively, and the final joint grade is shown in red at the bottom of the composite cutout. Each cutout is $\ang{;;9.9} \times \ang{;;9.9}$ in size.}
    \label{fig:B_Grade}
\end{figure*}

\begin{figure*}[!htb]%
    \centering
    \includegraphics[width=\textwidth]{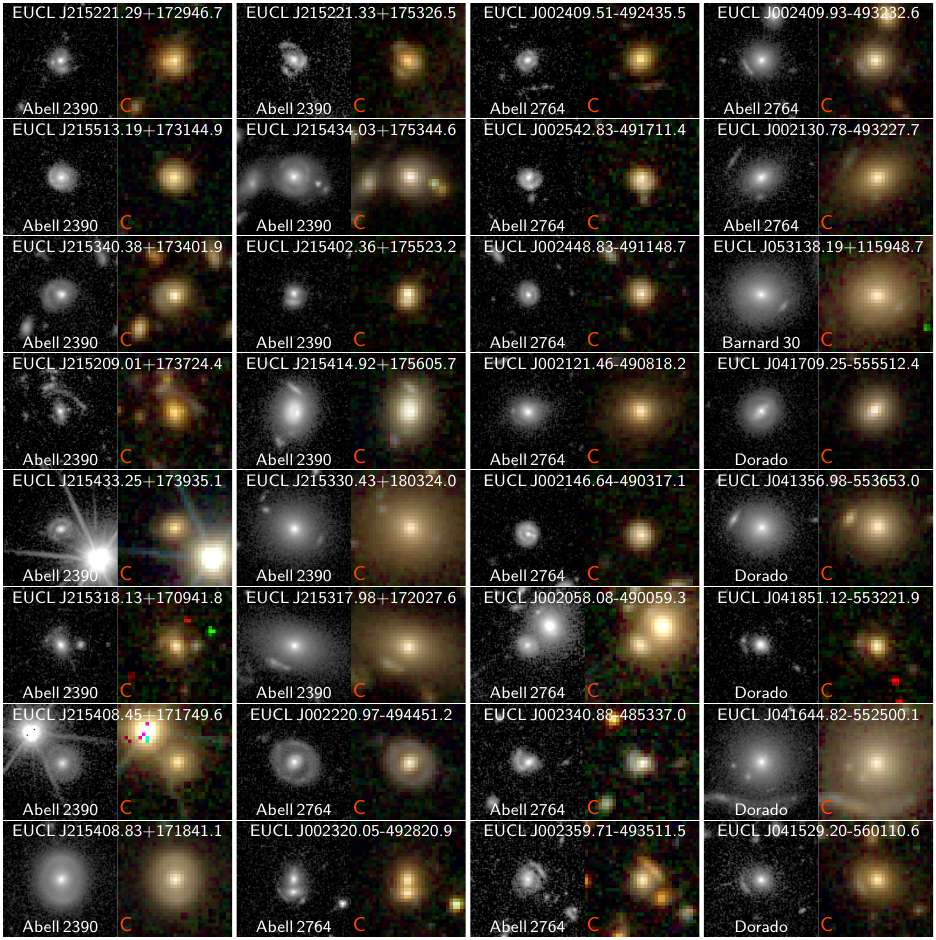}
    \caption{Mosaic of the 32 out of 52 grade C lens candidates from the second round of visual inspection. For each candidate, we show the high-resolution \IE-band cutout on the left and the lower-resolution \HE, \YE, \IE composite on the right. The IAU name and field name are displayed at the top and bottom of the \IE-band cutout, respectively, and the final joint grade is shown in red at the bottom of the composite cutout. Each cutout is $\ang{;;9.9} \times \ang{;;9.9}$ in size.}
    \label{fig:C_Grade_1}
\end{figure*}

\begin{figure*}[!htb]
    \centering
    \includegraphics[width=\textwidth]{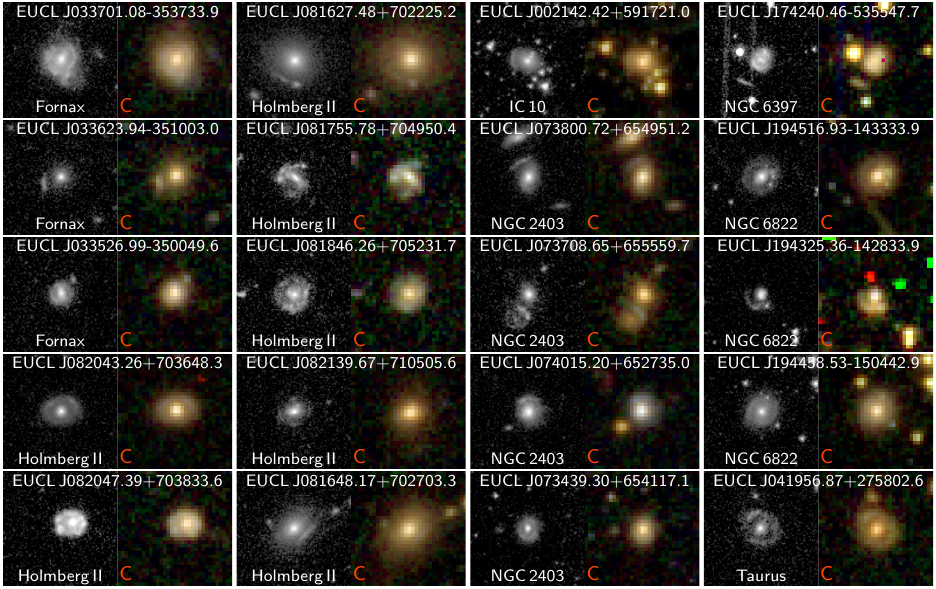}
    \caption{Mosaic of the remaining 20 grade C lens candidates from the second round of visual inspection. For each candidate, we show the high-resolution \IE-band cutout on the left and the lower-resolution \HE, \YE, \IE composite on the right. The IAU name and field name are displayed at the top and bottom of the \IE-band cutout, respectively, and the final joint grade is shown in red at the bottom of the composite cutout. Each cutout is $\ang{;;9.9} \times \ang{;;9.9}$ in size.}
    \label{fig:C_Grade_2}
\end{figure*}

  The 14 grade A candidates, representing definite lenses, are shown in \cref{fig:A_Grade}. The 31 grade B candidates, which exhibit strong lensing features but require further confirmation, are presented in \cref{fig:B_Grade}. Due to the larger number of grade C candidates, we have divided their presentation across \cref{fig:C_Grade_1,fig:C_Grade_2}, collectively displaying all 52 grade C objects that show potential lensing characteristics but necessitate additional investigation.

\subsection{Prediction}
We compare our findings with predictions made by  \cite{100000_lenses_3}, who estimated that the \Euclid mission would discover approximately \num{170000} strong gravitational lenses over its planned survey area of \num{15000} square degrees. Given that the 13 \Euclid ERO fields we analysed cover a total area of $8.12\deg^2$, we would naively expect $\sim$90 strong lenses in the ERO area, although this reduces to $\sim$70 after applying our $\IE < 23$ pre-selection. Our study has identified 97 grade A, B, and C lens candidates, a number that aligns closely with the theoretical expectations. However, experience suggests that most grade C candidates are unlikely to be lenses. Taking \cite{100000_lenses_3} as truth and pessimistically assuming only our grade A candidates are lenses, we would conclude that our search is $\sim$15\% complete. If the grade B candidates are also all lenses we have found $\sim$60 \% of all the lenses with $\IE<23$ that are in the ERO fields.

\section{Palomar spectroscopy}
\label{Sec:Spectroscopy}
We obtained optical spectroscopic follow-up of several of the grade A and grade B candidates using the Double Spectrograph (DBSP; \citealp{oke1982efficient}) on the 5m Hale telescope at Palomar Observatory between June and October 2024. \Cref{tab:spectroscopy} presents the targets for which we were able to measure at least one redshift in the candidate strong lens system. All nights had typical $\sim \ang{;;1.2}$ Palomar seeing. The June and July nights were photometric, while the October night had some clouds at the start, and end of the night, but was relatively clear at the time of the reported observation. For each observation, we obtained two or three exposures of $1200$ $\text{s}$ using the $\ang{;;1.5}$ slit, the 600 lines blue grating (blazed at 4000~\AA), the 5500~\AA\, dichroic, and the 316 lines red grating (blazed at 7500~\AA). The slits were aligned on the candidate lensing galaxy at a position angle to cover the putative lens feature. The data were reduced using standard techniques within the Image Reduction and Analysis Facility (IRAF).

\begin{figure*}[!htb]%
    \centering
    {{\includegraphics[width=1.\textwidth]{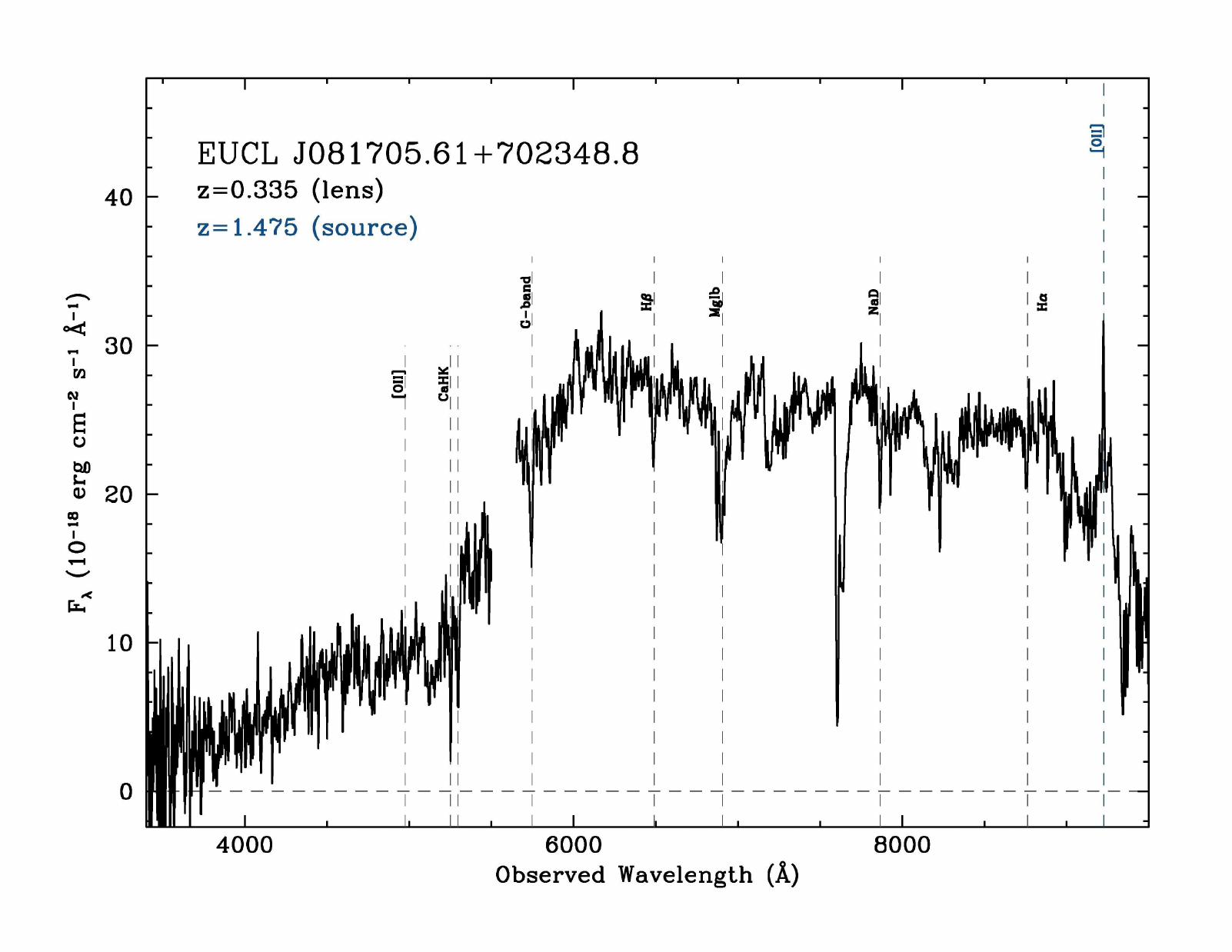}}}
    \caption{Spectrum of the grade A strong lens candidate EUCL\,J081705.61$+$702348.8 obtained at Palomar Observatory. The spectrum reveals an early-type galaxy at $z = 0.335$ with several characteristic absorption lines in addition to a strong 4000~\AA\, break. There is a strong additional line at 9222.8~\AA, which we associate with [\ion{O}{II}] emission from the lensed source galaxy at $z=1.475$.}
    \label{fig:Spectroscopy}
\end{figure*}

All the lensing galaxies proved to be early-type galaxies with Ca H \& K absorption and strong 4000~\AA\, breaks. One grade A system, EUCL\,J081705.61$+$702348.8, revealed redshifts for both the foreground lensing galaxy and the background source galaxy, where the latter had a slightly ($\sim \ang{;;1.7}$) offset emission line at 9222.8~\AA\, (\cref{fig:Spectroscopy}). If the emission feature was associated with the primary target galaxy at $z=0.335$, the implied rest-frame wavelength would be 6908~\AA, which does not correspond to any strong spectral features in galaxies, particularly early-type galaxies. Instead, the most plausible identification is [\ion{O}{II}] emission from a lensed star-forming galaxy at $z=1.475$.

\begin{figure*}
\centering
\includegraphics[width=0.24\textwidth]{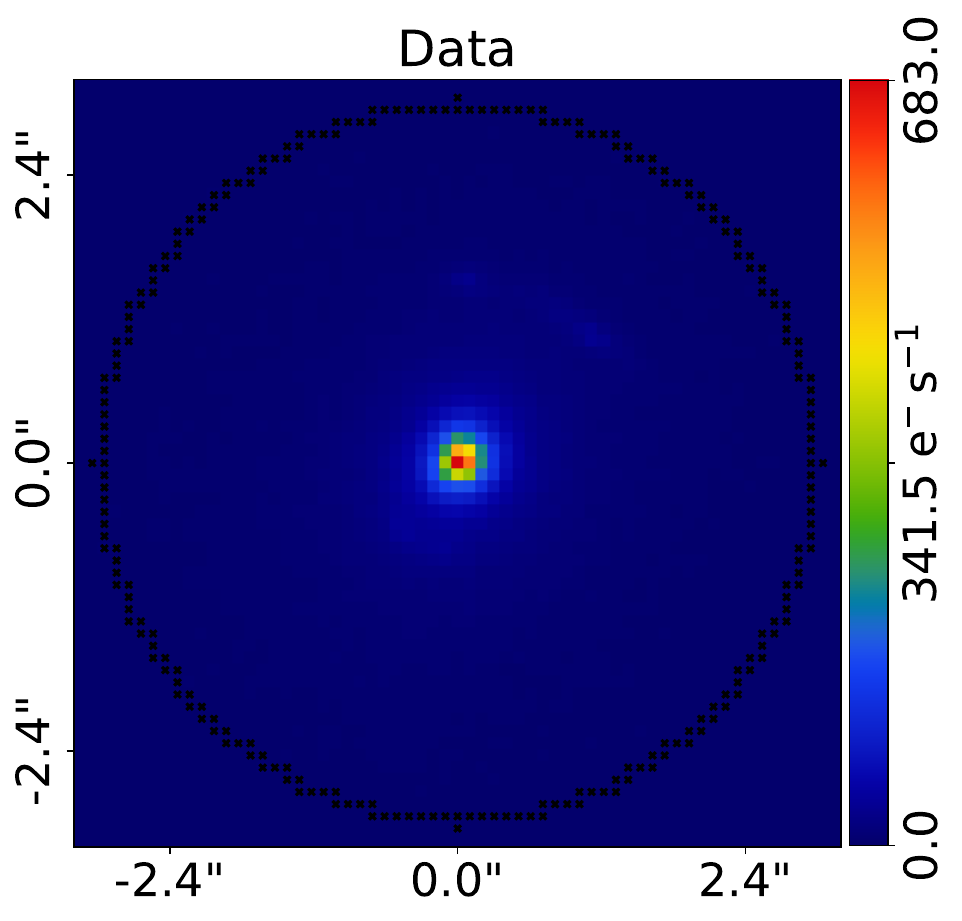}
\includegraphics[width=0.24\textwidth]{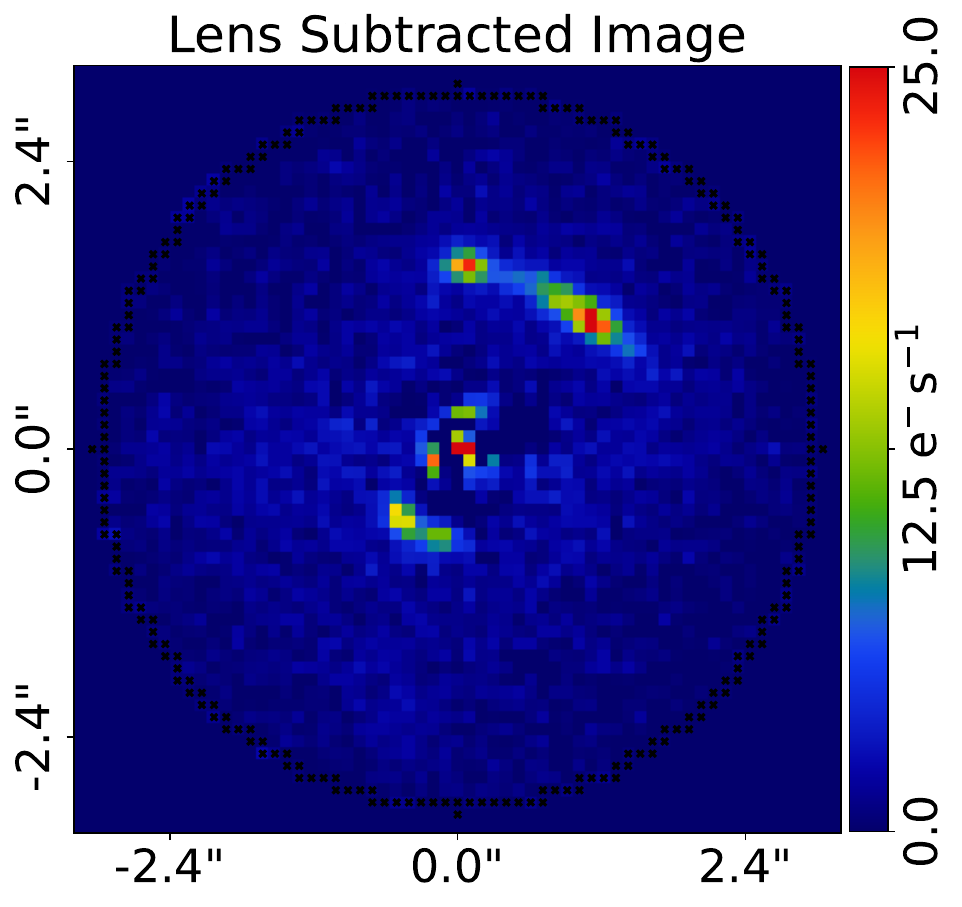}
\includegraphics[width=0.24\textwidth]{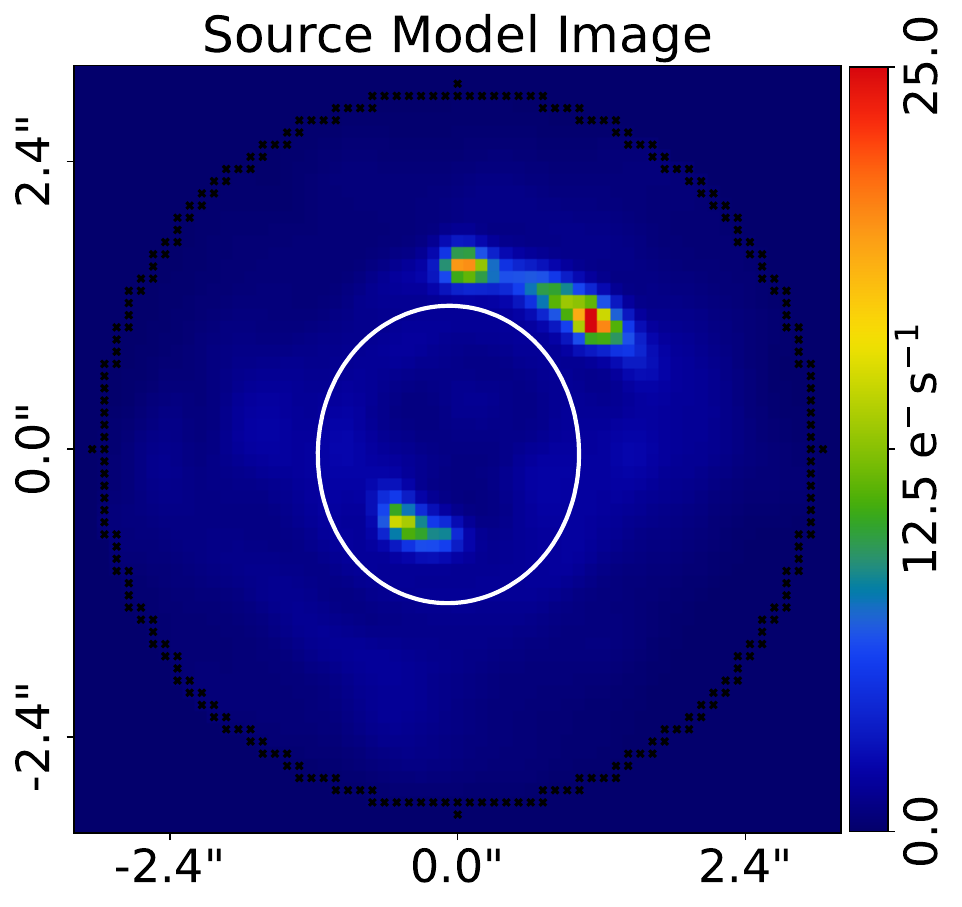}
\includegraphics[width=0.24\textwidth]{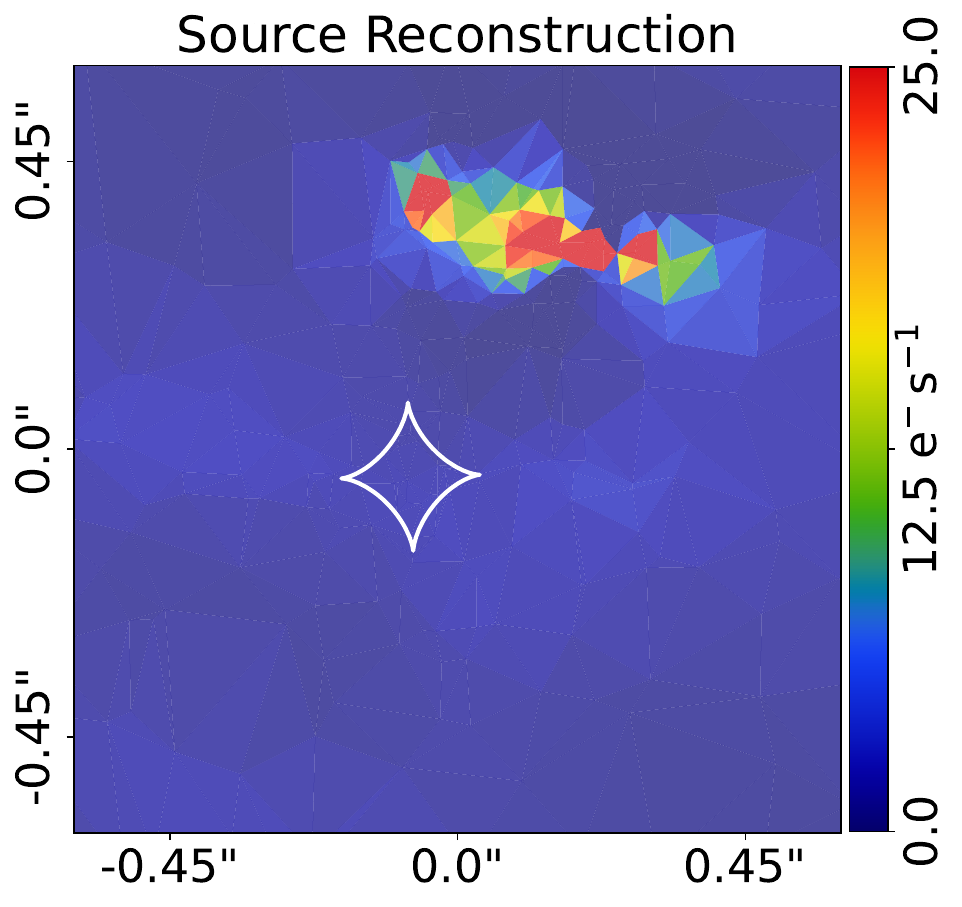}
\caption{
Lens model of the grade A system EUCL\,J081705.61+702348.8, created using the \Euclid strong lens modelling pipeline implemented with \texttt{PyAutoLens} \citep{Nightingale2018, Nightingale2021}. The panels show: (i) the observed VIS optical image; (ii) the lens-subtracted image; (iii) the lens model reconstruction of the lensed source; and (iv) the source galaxy's reconstructed unlensed emission in the source plane. The system is accurately modelled, providing strong evidence of its lensing nature. The mass model is a singular isothermal ellipsoid with external shear, the lens light is modelled using a multi-Gaussian expansion \citep{He2024}, and the source is reconstructed on an adaptive Delaunay mesh. The inferred Einstein radius is $\ang{;;1.18} \pm \ang{;;0.03}$.
}
\label{figure:LensModel}
\end{figure*}

To further confirm the lensing nature of EUCL\,J081705.61$+$702348.8, we modelled the system using the Euclid strong lens modelling pipeline implemented in \texttt{PyAutoLens} \citep{Nightingale2018, Nightingale2021}. A full description of the pipeline and application to a larger sample of lenses will be provided in the Q1 data release papers (Rojas et al. 2025, Walmsley et al. 2025). \Cref{figure:LensModel} presents the results for the VIS optical image. The foreground lens galaxy light is subtracted by fitting it with a multi-Gaussian expansion composed of 60 Gaussians \citep{He2024}. This reveals two distinct arcs in locations that are consistent with a singular isothermal ellipsoid (SIE) mass model configuration. \texttt{PyAutoLens} fits the SIE lens model, which the right panel of \Cref{figure:LensModel} shows accurately ray-traces both arcs to the same region of the source-plane. The source reconstruction reveals two separate doubly imaged emission components, which lie outside the tangential caustic but within the radial caustic, consistent with the tangential and radial critical curves predicted by our SIE mass model for this lens system. The SIE lens mass model includes an external shear, and the inferred Einstein radius is $\ang{;;1.18} \pm \ang{;;0.03}$.

\begin{table*}
\centering
\caption{Palomar spectroscopy of strong lens candidates.}
\begin{tabular*}{\textwidth}{@{\extracolsep{\fill}}lccccc@{}}
\hline
& & & & &\\[-9.5pt]
IAU Name & Grade & Obs Date (UT) & PA (deg) & $z$(lens) & $z$(source) \\
\hline
& & & & &\\[-9.5pt]
EUCL\,J081705.61$+$702348.8 & A & 2024 Oct 03 & $-$35 & 0.335 & 1.475 \\
EUCL\,J215252.97$+$171932.8 & B & 2024 Jul 10 & 60 & 0.418 & -- \\
EUCL\,J215405.95$+$181205.2 & A & 2024 Jun 02 & 114 & 0.474 & -- \\
EUCL\,J215447.40$+$174003.7 & A & 2024 Oct 03 & 10 & 0.716 & -- \\
\hline
\label{tab:spectroscopy}
\end{tabular*}
\end{table*}

\section{Discussion and conclusions}\label{sec:discussion}
We have presented a catalogue of strong gravitational lens candidates identified in the \Euclid Early Release Observations, covering 13 fields with a total area of $8.12\deg^2$. Our search methodology, combining multiple CNN models with visual classification, has yielded 97 lens candidates: 14 grade A, 31 grade B, and 52 grade C candidates. This multi-stage approach, using CNNs for initial candidate selection followed by careful visual inspection, has proven effective in producing a high-purity sample, particularly for grade A and B candidates.

The spectroscopic follow-up of our highest-quality candidates, notably EUCL\,J081705.61$+$702348.8, demonstrates the robustness of our classification system. Our spectroscopic analysis reveals an emission line at 9222.8~\AA\, that is most plausibly identified as [\ion{O}{II}] from a lensed star-forming galaxy at $z=1.475$, measuring the redshifts for both the foreground lensing galaxy ($z=0.335$) and the background source galaxy. The lens model implemented in \texttt{PyAutoLens} fits the two distinct arcs, revealing a source reconstruction with two doubly imaged emission components located outside the tangential caustic but within the radial caustic, thus confirming the strong lensing configuration. The lens modelling measured an Einstein radius of $\ang{;;1.18} \pm \ang{;;0.03}$.

We compare our findings with theoretical predictions, indicating that the discovery of 97 candidates (across all grades) in $8.12\deg^2$ is consistent with previous estimates of $\sim$70 strong lenses in this area (after applying our $\IE < 23$ pre-selection). Taking a conservative approach where only grade A candidates are considered confirmed lenses, our search achieves $\sim$15\% completeness, rising to $\sim$60\% if grade B candidates are included. Our lower completeness rate compared to Zoobot1 \citep{walmsley2023zoobot}, which identified 61 out of 68 known lenses (~90$\%$ completeness) in RP24, may be attributed to the fact that in our study the true number of lenses is unknown, making it more challenging than studies where the ground truth was known. Furthermore, our selection of only the top 650 candidates based on geometric mean scores for each field, while chosen to maintain a reasonable balance between completeness and false positive rate, may have excluded some genuine lens candidates that received lower CNN scores.

This work serves as a guide for future \Euclid lens searches, demonstrating that a broad initial CNN search followed by careful visual inspection is essential to build high-purity lens samples. Although our current analysis focuses on single-band ($\IE$) detections, the full potential of \Euclid lens searches will be realized through multi-band analysis, refined CNN architectures, and expanded spectroscopic follow-up campaigns. The lens modelling pipeline demonstrated here for EUCL\,J081705.61$+$702348.8 provides a template for future systematic analysis of larger lens samples in upcoming \Euclid data releases.

\begin{acknowledgements}
B.C.N. acknowledges support of DSSC and HPC cluster of RUG.
J.~A.~A.~B. and B.~C. acknowledge support from the Swiss National Science Foundation (SNSF).
C.T acknowledges the INAF grant 2022 LEMON. 
A.M.G. acknowledges the support of project PID2022-141915NB-C22 funded by MCIU/AEI/10.13039/501100011033 and FEDER/UE. Based on observations obtained at the Hale Telescope, Palomar Observatory, as part of a collaborative agreement between the Caltech Optical Observatories and the Jet Propulsion Laboratory.
This work made use of \texttt{Astropy}: a community-developed core Python package and an ecosystem of tools and resources for astronomy \citep{astropy13,astropy18,astropy22}, \texttt{NumPy} \citep{numpy}, \texttt{Matplotlib} \citep{matplotlib}, and \texttt{pandas} \citep{mckinney2011pandas}.
\AckERO
\AckEC
\end{acknowledgements}

%
%

\bibliography{Euclid,Not_Euclid}
\label{LastPage}
\begin{appendix}
  \onecolumn 
\section{Lens candidates\label{apdx:A}}
We show the details of 97 lens candidates which are distributed as 14 grade A candidates, 31 grade B candidates, and 52 grade C candidates.



\setlength\LTleft{0pt}
\setlength\LTright{0pt}
\setlength\LTcapwidth{\linewidth}
\begin{longtable}{@{\extracolsep{\fill}}ccr@{.}lr@{.}lcccccccc@{}}

\caption{Probable lens candidates discovered with CNN classifiers. ID, IAU Name, Fields, RA, DEC, Classification prediction scores for five different networks (P1, P2, P3, P4 and P5), Geometric Mean (GM) of the five prediction scores and visual inspection grade are shown. Here, P1 refers to the \texttt{Lens-CLR} network, P2 refers to the \texttt{Denselens}, P3 refers to the \texttt{4-Layer CNN}, P4 refers to the \texttt{Naberrie} and P5 refers to the \texttt{MRC95}.}\\
\hline
IAU Name & Field & \multicolumn{2}{c}{RA} & \multicolumn{2}{c}{Dec} & P1 & P2 & P3 & P4 & P5 & GM & Grade \\
\hline

EUCL\,J215447.40+174003.7 & Abell\,2390 & 328&697535 & 17&667696 & 0.990 & 0.880 & 0.936 & 0.006 & 0.999 & 0.353 & A \\
EUCL\,J215405.95+181205.2 & Abell\,2390 & 328&524807 & 18&201453 & 0.966 & 0.999 & 0.974 & 0.704 & 0.998 & 0.920 & A \\
EUCL\,J002251.41$$-$$494843.9 & Abell\,2764 & 5&714233 & -49&812211 & 0.211 & 0.470 & 0.998 & 0.093 & 0.999 & 0.392 & A \\
EUCL\,J002539.83$$-$$491515.5 & Abell\,2764 & 6&415961 & -49&254323 & 0.982 & 0.817 & 0.822 & 0.172 & 0.999 & 0.647 & A \\
EUCL\,J002330.90$$-$$485506.4 & Abell\,2764 & 5&878780 & -48&918448 & 0.972 & 0.980 & 0.982 & 0.503 & 0.999 & 0.860 & A \\
EUCL\,J053110.94+121542.8 & Barnard\,30 & 82&795595 & 12&261915 & 0.750 & 0.014 & 0.989 & 0.854 & 0.786 & 0.373 & A \\
EUCL\,J041819.40$$-$$554659.7 & Dorado & 64&580860 & -55&783276 & 0.981 & 0.693 & 0.881 & 0.382 & 1.000 & 0.744 & A \\
EUCL\,J041729.09$$-$$554120.0 & Dorado & 64&371238 & -55&688895 & 0.889 & 0.572 & 1.000 & 0.471 & 1.000 & 0.751 & A \\
EUCL\,J033706.49$$-$$351842.4 & Fornax & 54&277074 & -35&311798 & 0.767 & 0.235 & 0.999 & 0.315 & 0.999 & 0.563 & A \\
EUCL\,J033646.89$$-$$351138.8 & Fornax & 54&195391 & -35&194112 & 0.994 & 0.995 & 0.991 & 0.014 & 0.999 & 0.428 & A \\
EUCL\,J081945.02+702002.2 & Holmberg\,II & 124&937617 & 70&333964 & 0.024 & 0.432 & 0.422 & 0.008 & 0.999 & 0.131 & A \\
EUCL\,J081705.61+702348.8 & Holmberg\,II & 124&273402 & 70&396896 & 0.930 & 0.946 & 0.992 & 0.891 & 0.999 & 0.951 & A \\
EUCL\,J002035.05+585556.2 & IC\,10 & 5&146066 & 58&932298 & 0.131 & 0.801 & 0.858 & 0.763 & 0.999 & 0.585 & A \\
EUCL\,J191245.11$$-$$634926.2 & NGC\,6744 & 288&187969 & -63&823954 & 0.413 & 0.738 & 0.864 & 0.901 & 0.999 & 0.750 & A \\
EUCL\,J215439.38+173354.6 & Abell\,2390 & 328&664120 & 17&565190 & 0.950 & 0.130 & 0.999 & 0.177 & 0.999 & 0.465 & B \\
EUCL\,J215354.10+175419.0 & Abell\,2390 & 328&475438 & 17&905296 & 0.931 & 0.236 & 0.987 & 0.928 & 0.826 & 0.698 & B \\
EUCL\,J215252.97+171932.8 & Abell\,2390 & 328&220730 & 17&325797 & 0.977 & 0.691 & 0.997 & 0.862 & 0.508 & 0.783 & B \\
EUCL\,J002136.55$$-$$492944.5 & Abell\,2764 & 5&402328 & -49&495699 & 0.993 & 0.917 & 0.791 & 0.097 & 0.940 & 0.581 & B \\
EUCL\,J002313.02$$-$$492640.7 & Abell\,2764 & 5&804288 & -49&444665 & 0.959 & 0.927 & 0.998 & 0.215 & 0.999 & 0.718 & B \\
EUCL\,J002352.59$$-$$492403.3 & Abell\,2764 & 5&969141 & -49&400920 & 0.104 & 0.735 & 0.999 & 0.073 & 0.999 & 0.354 & B \\
EUCL\,J002255.86$$-$$492228.2 & Abell\,2764 & 5&732785 & -49&374505 & 0.019 & 0.535 & 0.800 & 0.576 & 0.999 & 0.342 & B \\
EUCL\,J002304.81$$-$$491845.7 & Abell\,2764 & 5&770044 & -49&312716 & 0.689 & 0.474 & 0.494 & 0.511 & 0.996 & 0.607 & B \\
EUCL\,J002442.16$$-$$491345.8 & Abell\,2764 & 6&175704 & -49&229391 & 0.941 & 0.738 & 0.121 & 0.637 & 0.931 & 0.549 & B \\
EUCL\,J002148.37$$-$$490541.5 & Abell\,2764 & 5&451552 & -49&094878 & 0.996 & 0.798 & 0.999 & 0.386 & 0.999 & 0.789 & B \\
EUCL\,J002131.51$$-$$490533.6 & Abell\,2764 & 5&381318 & -49&092670 & 0.892 & 0.483 & 0.998 & 0.774 & 0.111 & 0.517 & B \\
EUCL\,J002133.06$$-$$490020.3 & Abell\,2764 & 5&387787 & -49&005661 & 0.329 & 0.464 & 0.999 & 0.050 & 0.999 & 0.377 & B \\
EUCL\,J041425.56$$-$$555929.9 & Dorado & 63&606523 & -55&991655 & 0.993 & 0.216 & 1.000 & 0.894 & 0.999 & 0.719 & B \\
EUCL\,J041841.96$$-$$555503.4 & Dorado & 64&674870 & -55&917617 & 0.962 & 0.068 & 0.519 & 0.909 & 0.816 & 0.517 & B \\
EUCL\,J041556.54$$-$$555447.6 & Dorado & 63&985615 & -55&913242 & 0.912 & 0.468 & 0.997 & 0.845 & 0.886 & 0.796 & B \\
EUCL\,J041629.29$$-$$553257.7 & Dorado & 64&122045 & -55&549366 & 0.894 & 0.974 & 0.999 & 0.804 & 0.635 & 0.850 & B \\
EUCL\,J041335.15$$-$$552856.9 & Dorado & 63&396468 & -55&482498 & 0.999 & 0.422 & 0.863 & 0.827 & 0.955 & 0.779 & B \\
EUCL\,J081943.34+702335.6 & Holmberg\,II & 124&930595 & 70&393246 & 0.345 & 0.293 & 0.998 & 0.785 & 0.999 & 0.602 & B \\
EUCL\,J081959.64+710219.7 & Holmberg\,II & 124&998533 & 71&038816 & 0.755 & 0.493 & 1.000 & 0.074 & 0.999 & 0.487 & B \\
EUCL\,J001727.46+591801.9 & IC\,10 & 4&364447 & 59&300532 & 0.400 & 0.339 & 0.948 & 0.898 & 0.997 & 0.649 & B \\
EUCL\,J054527.40$$-$$000548.7 & Messier\,78 & 86&364175 & -0&096864 & 0.232 & 0.355 & 0.463 & 0.785 & 0.001 & 0.124 & B \\
EUCL\,J054807.82+001045.5 & Messier\,78 & 87&032612 & 0&179325 & 0.759 & 0.352 & 0.059 & 0.071 & 0.999 & 0.258 & B \\
EUCL\,J074006.13+655047.9 & NGC\,2403 & 115&025578 & 65&846644 & 0.574 & 0.304 & 0.303 & 0.917 & 0.996 & 0.546 & B \\
EUCL\,J073450.92+655829.7 & NGC\,2403 & 113&712202 & 65&974936 & 0.246 & 0.104 & 0.562 & 0.885 & 0.999 & 0.418 & B \\
EUCL\,J073851.49+655650.2 & NGC\,2403 & 114&714583 & 65&947284 & 0.992 & 0.723 & 0.977 & 0.560 & 0.949 & 0.821 & B \\
EUCL\,J073637.17+660131.1 & NGC\,2403 & 114&154879 & 66&025315 & 0.556 & 0.157 & 0.996 & 0.754 & 0.692 & 0.539 & B \\
EUCL\,J073600.54+651440.8 & NGC\,2403 & 114&002254 & 65&244693 & 0.985 & 0.612 & 0.999 & 0.728 & 0.788 & 0.809 & B \\
EUCL\,J073924.78+651814.3 & NGC\,2403 & 114&853255 & 65&303999 & 0.766 & 0.367 & 0.580 & 0.051 & 0.999 & 0.384 & B \\
EUCL\,J073609.14+651905.3 & NGC\,2403 & 114&038123 & 65&318153 & 0.702 & 0.925 & 0.999 & 0.463 & 0.742 & 0.741 & B \\
EUCL\,J073910.33+652326.4 & NGC\,2403 & 114&793076 & 65&390688 & 0.982 & 0.759 & 0.999 & 0.254 & 0.999 & 0.717 & B \\
EUCL\,J190644.53$$-$$635426.5 & NGC\,6744 & 286&685558 & -63&907383 & 0.169 & 0.511 & 0.998 & 0.009 & 0.999 & 0.241 & B \\
EUCL\,J215221.29+172946.7 & Abell\,2390 & 328&088745 & 17&496325 & 0.872 & 0.404 & 0.987 & 0.104 & 0.999 & 0.515 & C \\
EUCL\,J215513.19+173144.9 & Abell\,2390 & 328&804966 & 17&529147 & 0.344 & 0.392 & 0.900 & 0.392 & 0.999 & 0.544 & C \\
EUCL\,J215340.38+173401.9 & Abell\,2390 & 328&418271 & 17&567201 & 0.564 & 0.673 & 0.943 & 0.770 & 0.999 & 0.773 & C \\
EUCL\,J215209.01+173724.4 & Abell\,2390 & 328&037552 & 17&623459 & 0.319 & 0.316 & 0.895 & 0.271 & 0.999 & 0.476 & C \\
EUCL\,J215433.25+173935.1 & Abell\,2390 & 328&638549 & 17&659777 & 0.984 & 0.137 & 0.304 & 0.952 & 0.999 & 0.522 & C \\
EUCL\,J215318.13+170941.8 & Abell\,2390 & 328&325546 & 17&161622 & 0.970 & 0.298 & 0.956 & 0.406 & 0.866 & 0.627 & C \\
EUCL\,J215408.45+171749.6 & Abell\,2390 & 328&535221 & 17&297134 & 0.984 & 0.091 & 0.403 & 0.949 & 0.999 & 0.509 & C \\
EUCL\,J215408.83+171841.1 & Abell\,2390 & 328&536832 & 17&311437 & 0.002 & 0.327 & 0.162 & 0.880 & 0.999 & 0.165 & C \\
EUCL\,J215221.33+175326.5 & Abell\,2390 & 328&088883 & 17&890708 & 0.976 & 0.723 & 0.847 & 0.525 & 0.999 & 0.793 & C \\
EUCL\,J215434.03+175344.6 & Abell\,2390 & 328&641815 & 17&895743 & 0.962 & 0.085 & 0.997 & 0.944 & 0.999 & 0.598 & C \\
EUCL\,J215402.36+175523.2 & Abell\,2390 & 328&509837 & 17&923113 & 0.776 & 0.699 & 0.985 & 0.849 & 0.999 & 0.854 & C \\
EUCL\,J215414.92+175605.7 & Abell\,2390 & 328&562193 & 17&934936 & 0.929 & 0.834 & 0.990 & 0.172 & 0.999 & 0.667 & C \\
EUCL\,J215330.43+180324.0 & Abell\,2390 & 328&376812 & 18&056668 & 0.925 & 0.069 & 0.985 & 0.919 & 0.999 & 0.566 & C \\
EUCL\,J215317.98+172027.6 & Abell\,2390 & 328&324921 & 17&341027 & 0.835 & 0.758 & 0.997 & 0.903 & 0.774 & 0.849 & C \\
EUCL\,J002220.97$$-$$494451.2 & Abell\,2764 & 5&587376 & -49&747565 & 0.154 & 0.553 & 0.999 & 0.023 & 0.999 & 0.288 & C \\
EUCL\,J002320.05$$-$$492820.9 & Abell\,2764 & 5&833578 & -49&472500 & 0.977 & 0.126 & 0.981 & 0.802 & 0.999 & 0.627 & C \\
EUCL\,J002409.51$$-$$492435.5 & Abell\,2764 & 6&039630 & -49&409882 & 0.964 & 0.519 & 0.999 & 0.415 & 0.999 & 0.730 & C \\
EUCL\,J002542.83$$-$$491711.4 & Abell\,2764 & 6&428471 & -49&286524 & 0.701 & 0.178 & 0.943 & 0.504 & 0.999 & 0.568 & C \\
EUCL\,J002448.83$$-$$491148.7 & Abell\,2764 & 6&203498 & -49&196888 & 0.225 & 0.907 & 0.999 & 0.145 & 0.999 & 0.495 & C \\
EUCL\,J002121.46$$-$$490818.2 & Abell\,2764 & 5&339445 & -49&138403 & 0.943 & 0.407 & 0.818 & 0.798 & 0.999 & 0.758 & C \\
EUCL\,J002146.64$$-$$490317.1 & Abell\,2764 & 5&444347 & -49&054750 & 0.646 & 0.653 & 0.999 & 0.066 & 0.999 & 0.489 & C \\
EUCL\,J002058.08$$-$$490059.3 & Abell\,2764 & 5&242019 & -49&016492 & 0.979 & 0.057 & 0.674 & 0.909 & 0.999 & 0.510 & C \\
EUCL\,J002340.88$$-$$485337.0 & Abell\,2764 & 5&920356 & -48&893632 & 0.422 & 0.643 & 0.991 & 0.156 & 0.999 & 0.530 & C \\
EUCL\,J002359.71$$-$$493511.5 & Abell\,2764 & 5&998816 & -49&586542 & 0.882 & 0.325 & 0.999 & 0.112 & 0.999 & 0.503 & C \\
EUCL\,J002409.93$$-$$493232.6 & Abell\,2764 & 6&041410 & -49&542403 & 0.960 & 0.690 & 0.999 & 0.832 & 0.998 & 0.887 & C \\
EUCL\,J002130.78$$-$$493227.7 & Abell\,2764 & 5&378283 & -49&541048 & 0.987 & 0.362 & 0.280 & 0.687 & 0.786 & 0.558 & C \\
EUCL\,J053138.19+115948.7 & Barnard\,30 & 82&909158 & 11&996874 & 0.917 & 0.253 & 0.606 & 0.914 & 0.913 & 0.651 & C \\
EUCL\,J041709.25$$-$$555512.4 & Dorado & 64&288546 & -55&920135 & 0.955 & 0.002 & 0.946 & 0.922 & 1.000 & 0.608 & C \\
EUCL\,J041356.98$$-$$553653.0 & Dorado & 63&487439 & -55&614728 & 0.553 & 0.523 & 0.995 & 0.886 & 1.000 & 0.761 & C \\
EUCL\,J041851.12$$-$$553221.9 & Dorado & 64&713024 & -55&539426 & 0.597 & 0.039 & 0.991 & 0.401 & 0.788 & 0.451 & C \\
EUCL\,J041644.82$$-$$552500.1 & Dorado & 64&186769 & -55&416712 & 0.927 & 0.260 & 0.863 & 0.622 & 0.565 & 0.593 & C \\
EUCL\,J041529.20$$-$$560110.6 & Dorado & 63&871674 & -56&019634 & 0.958 & 0.759 & 0.872 & 0.213 & 0.996 & 0.670 & C \\
EUCL\,J033701.08$$-$$353733.9 & Fornax & 54&254540 & -35&626088 & 0.411 & 0.485 & 0.998 & 0.061 & 0.999 & 0.414 & C \\
EUCL\,J033623.94$$-$$351003.0 & Fornax & 54&099781 & -35&167520 & 0.887 & 0.194 & 0.259 & 0.496 & 0.995 & 0.466 & C \\
EUCL\,J033526.99$$-$$350049.6 & Fornax & 53&862486 & -35&013784 & 0.638 & 0.146 & 0.156 & 0.297 & 0.999 & 0.336 & C \\
EUCL\,J082043.26+703648.3 & Holmberg\,II & 125&180250 & 70&613437 & 0.994 & 0.394 & 0.849 & 0.807 & 0.999 & 0.769 & C \\
EUCL\,J082047.39+703833.6 & Holmberg\,II & 125&197494 & 70&642678 & 0.555 & 0.837 & 0.999 & 0.233 & 0.999 & 0.641 & C \\
EUCL\,J081627.48+702225.2 & Holmberg\,II & 124&114541 & 70&373679 & 0.920 & 0.652 & 0.998 & 0.919 & 0.982 & 0.884 & C \\
EUCL\,J081755.78+704950.4 & Holmberg\,II & 124&482442 & 70&830679 & 0.912 & 0.027 & 0.999 & 0.532 & 0.999 & 0.420 & C \\
EUCL\,J081846.26+705231.7 & Holmberg\,II & 124&692781 & 70&875493 & 0.971 & 0.650 & 0.998 & 0.267 & 0.998 & 0.700 & C \\
EUCL\,J082139.67+710505.6 & Holmberg\,II & 125&415301 & 71&084913 & 0.618 & 0.274 & 0.997 & 0.164 & 0.999 & 0.488 & C \\
EUCL\,J081648.17+702703.3 & Holmberg\,II & 124&200716 & 70&450925 & 0.981 & 0.349 & 0.771 & 0.155 & 0.995 & 0.527 & C \\
EUCL\,J002142.42+591721.0 & IC\,10 & 5&426777 & 59&289174 & 0.720 & 0.163 & 0.993 & 0.886 & 0.999 & 0.635 & C \\
EUCL\,J073800.72+654951.2 & NGC\,2403 & 114&503007 & 65&830916 & 0.906 & 0.668 & 0.995 & 0.242 & 0.999 & 0.680 & C \\
EUCL\,J073708.65+655559.7 & NGC\,2403 & 114&286066 & 65&933256 & 0.790 & 0.092 & 0.482 & 0.572 & 0.999 & 0.458 & C \\
EUCL\,J074015.20+652735.0 & NGC\,2403 & 115&063350 & 65&459746 & 0.238 & 0.929 & 0.999 & 0.354 & 0.998 & 0.601 & C \\
EUCL\,J073439.30+654117.1 & NGC\,2403 & 113&663751 & 65&688102 & 0.081 & 0.898 & 0.943 & 0.013 & 0.999 & 0.248 & C \\
EUCL\,J174240.46$$-$$535547.7 & NGC\,6397 & 265&668618 & -53&929936 & 0.440 & 0.298 & 0.743 & 0.929 & 0.999 & 0.618 & C \\
EUCL\,J194516.93$$-$$143333.9 & NGC\,6822 & 296&320580 & -14&559422 & 0.972 & 0.027 & 1.000 & 0.078 & 0.999 & 0.376 & C \\

EUCL\,J194325.36$$-$$142833.9 & NGC\,6822 & 295&855690 & $$-$$14&476105 & 0.323 & 0.294 & 0.779 & 0.506 & 1.000 & 0.518 & C \\
EUCL\,J194458.53$$-$$150442.9 & NGC\,6822 & 296&243914 & $$-$$15&078600 & 0.730 & 0.094 & 0.845 & 0.918 & 1.000 & 0.563 & C \\
EUCL\,J041956.87+275802.6 & Taurus & 64&986995 & 27&967397 & 0.650 & 0.506 & 0.609 & 0.302 & 0.997 & 0.570 & C \\

\bottomrule
\label{sec:table_catalog}
\end{longtable}


\FloatBarrier



\newpage
\section{Summary of the ERO fields}
We present here the summary of the 16 \Euclid ERO fields (except Perseus). The \Euclid VIS instrument provides high-quality optical imaging through a single broad optical band (550-900 nm). The VIS camera features a pixel scale of \ang{;;0.1} per pixel. The point spread function (PSF) of VIS has a full width at half maximum (FWHM) of $\lesssim \ang{;;0.18}$ across the entire field of view \citep{cropper20169}. For the ERO fields, each pointing consists of four dithered exposures of 565 seconds each, resulting in a total exposure time of 2260 seconds. This yields a typical $5\sigma$ point-source depth of $\IE \approx 26.2$ mag \citep{Scaramella-EP1}. The VIS images demonstrate excellent depth and resolution, with a median signal-to-noise ratio (S/N) of $\sim$10$\sigma$ for objects at $\IE = 24.5$ mag. 
\begin{table}[htbp]
\caption{Summary of all the 16 \Euclid ERO Fields (except Perseus) including their corresponding area, and their corresponding catalogue size, which represents the number of sources in each field.}
\begin{tabular}{c c c c c|}
\hline
\\[-10px]
Field name  & Area ($\deg^2$) & Catalogue size \\ 
\hline
\\[-9px]
Abell\,2390  & $0.75$ & \phantom{1\,}\num{469056} \\ 

Abell\,2764    & $0.75$ & \phantom{1\,}\num{542729} \\ 
Barnard\,30    & $0.6$\phantom{5} & \phantom{1\,}\num{163063} \\ 
Dorado        & $0.6$\phantom{5} & \phantom{1\,}\num{518445} \\ 
Fornax       & $0.57$ & \phantom{1\,}\num{369315} \\ 
Holmberg\,II  & $0.6$\phantom{5} & \phantom{1\,}\num{466276} \\ 
Horsehead Nebula  & $0.58$ & \phantom{1\,}\num{157264} \\ 
IC\,10        & $0.62$ & \num{1403807} \\ 
IC\,342       & $0.59$ & \num{2033293} \\ 
M78          & $0.6$\phantom{5} & \phantom{1\,}\num{116373} \\ 
NGC\,2403     & $0.6$\phantom{5} & \num{1152966} \\ 
NGC\,6254     & $0.6$\phantom{5} & \phantom{1\,}\num{413297} \\ 
NGC\,6397     & $0.61$ & \phantom{1\,}\num{782612} \\ 
NGC\,6744     & $0.6$\phantom{5} & \phantom{1\,}\num{924913} \\ 
NGC\,6822     & $0.6$\phantom{5} & \num{1694021} \\ 
Taurus       & $0.61$ & \phantom{1\,}\num{123726} \\ 
\hline
\end{tabular}
\label{table:ero_statistics}
\end{table}


\end{appendix}


\end{document}